\documentclass[letterpaper,aps,twocolumn,showpacs,floats,prb]{revtex4}

\usepackage{amsmath}
\usepackage{graphicx,epsfig,dsfont}
\usepackage{amssymb}

\newcommand{\nn}{\nonumber}

\providecommand{\vect}[1]{\boldsymbol{#1}}

\begin{document}

\title{
Magnon-skyrmion scattering in chiral magnets
}

\author{Christoph Sch\"utte}
\author{Markus Garst}
\affiliation{Institut f\"ur Theoretische Physik, Universit\"at zu K\"oln,
Z\"ulpicher Str. 77, 50937 K\"oln, Germany
}

\begin{abstract}
Chiral magnets support topological skyrmion textures due to the Dzyaloshinskii-Moriya spin-orbit interaction. In the presence of a sufficiently large applied magnetic field, such skyrmions are large amplitude excitations of the field-polarized magnetic state. We investigate analytically the interaction between such a skyrmion excitation and its small amplitude fluctuations, i.e., the magnons in a clean two-dimensional chiral magnet. 
The magnon spectrum is found to include two magnon-skyrmion bound states corresponding to a breathing mode and, for intermediate fields, a quadrupolar mode, which will give rise to subgap magnetic and electric resonances.
Due to the skyrmion topology, the magnons scatter from an Aharonov-Bohm flux density that leads to skew and rainbow scattering, characterized by an 
asymmetric differential cross section with, in general, multiple peaks.
As a consequence of the skew scattering, a finite density of skyrmions will generate a topological magnon Hall effect.
Using the conservation law for the energy-momentum tensor, we demonstrate that the magnons also transfer momentum to the skyrmion. As a consequence, a magnon current leads to magnon pressure reflected in a momentum-transfer force in the Thiele equation of motion for the skyrmion. 
This force is reactive and governed by the scattering cross sections of the skyrmion;
it causes not only a finite skyrmion velocity but also a large skyrmion Hall effect.
Our results provide, in particular, the basis for a theory of skyrmion caloritronics 
for a dilute skyrmion gas in clean insulating chiral magnets. 
\end{abstract}

\date{\today}


\maketitle

\section{Introduction} 
\label{sec:Introduction}

\begin{figure}
\centering
\includegraphics[width=0.8\columnwidth]{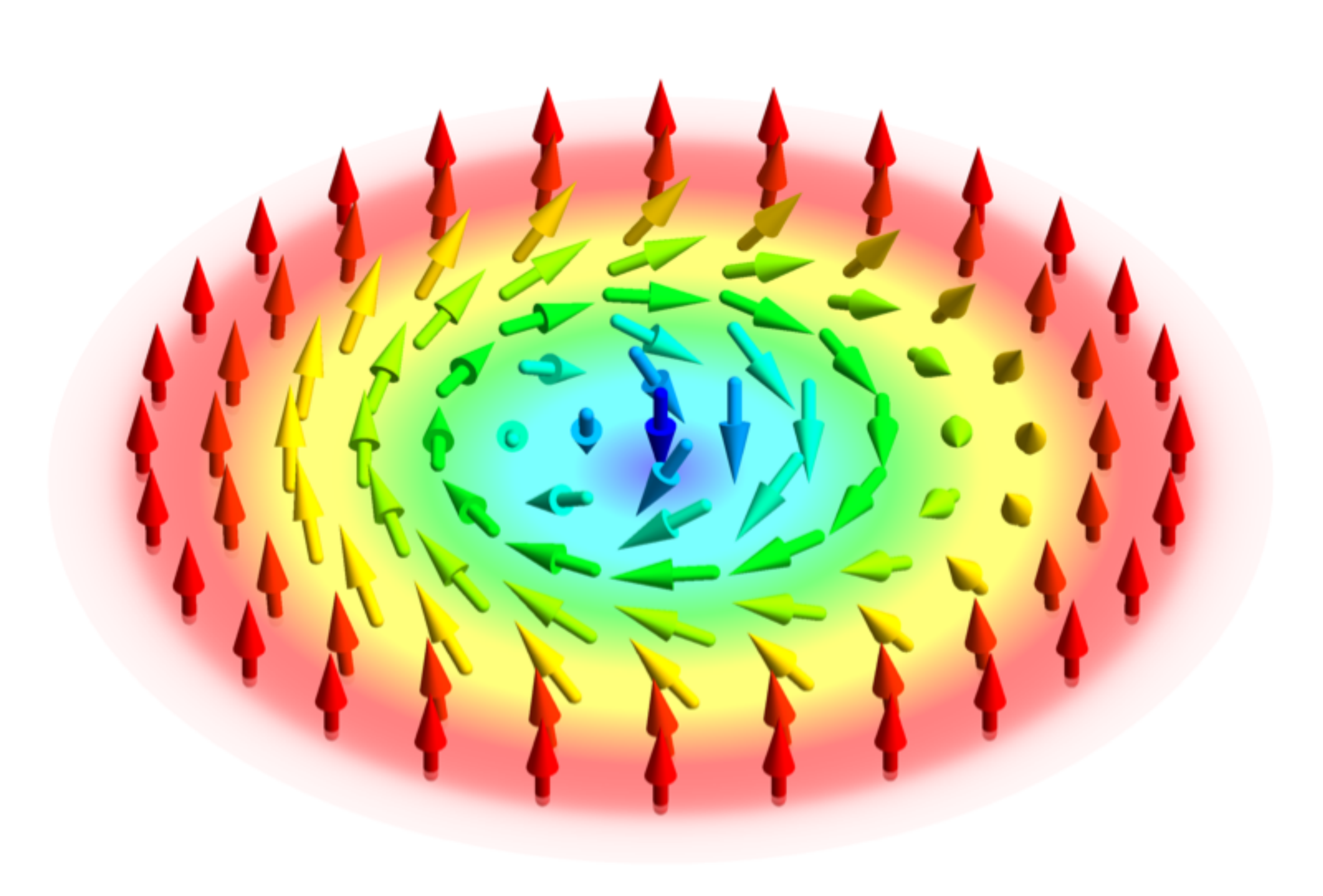}
\caption{Skyrmion texture of a chiral magnet.}
\label{fig:skyrmionTexture}
\end{figure}

The discovery of a magnetic skyrmion lattice in the cubic chiral magnets by M\"uhlbauer {\it et al}.\cite{Muehlbauer2009} has triggered a flurry of interest in magnetic skyrmion textures. These magnets inherit the chirality from their B20 chiral atomic crystal structure allowing for a spin-orbit Dzyaloshinskii-Moriya interaction, that energetically stabilizes twisted modulated magnetic textures like helices and skyrmions, see Fig.~\ref{fig:skyrmionTexture}. The possibility of magnetic skyrmion textures in the B20 compounds has been envisioned in early seminal work by Bogdanov and collaborators.\cite{Bogdanov1989,Boganov1994,Roessler2006} They encompass a variety of materials with different electronic characteristics, for example, the metals MnSi\cite{Muehlbauer2009,Adams2011} and FeGe,\cite{Yu2011} the semiconductor Fe$_{1-x}$Co$_x$Si,\cite{Muenzer2010,Yu2010} and the insulator Cu$_2$OSeO$_3$,\cite{Seki2012,Adams2012} that nevertheless share the same magnetic properties and possess similar magnetic phase diagrams.

The excitement aroused by skyrmions is attributed to their topological properties. They are characterized by a 
finite topological winding number that is, e.g., at the origin of a topological Hall effect,\cite{Neubauer2009,Lee2009} a skyrmion-flow Hall effect,\cite{Schulz2012} and a concomitant emergent electrodynamics\cite{Volovik2003,Nagaosa2012} in the metallic B20 compounds. 
Moreover, their topological origin also results in a finite gyrocoupling vector in the Thiele equation,\cite{Thiele1973} that describes their magnetization dynamics, and, as a consequence, the skyrmion motion is governed by a strong spin-Magnus force.\cite{Stone1996,Jonietz2010,Everschor2011,Everschor2012}
This peculiar dynamics combined with the smoothness of the skyrmion texture allows for spin-transfer torque phenomena at ultralow threshold currents,\cite{Jonietz2010,Yu2012,Iwasaki2013,Lin2013-1,Lin2013-2} which makes magnetic skyrmion matter interesting for spintronic applications;\cite{Sampaio2013} see Ref.~\onlinecite{Nagaosa2013} for a recent review.
In insulating chiral magnets, skyrmions are also associated with interesting thermal spin-transport effects. A thermal gradient is predicted to induce a skyrmion motion that, counterintuitively, is towards the heat source together with a skyrmion Hall effect.\cite{Kong2013,Lin2013-3,Kovalev2014}
Moreover, a thermal skyrmion ratchet has been realized with the help of the topological magnon Hall effect that arises due to magnon skew scattering off skyrmions in the material.\cite{Mochizuki2014} 

A prerequisite for a better understanding of such phenomena, however, is a detailed analysis of the magnon-skyrmion interaction. 
For the skyrmion lattice, it is known that apart from the long-wavelength spin-wave excitations\cite{Petrova2011,Zang2011} there are three magnetic resonances with finite excitation frequencies, a single breathing mode and two gyration modes.\cite{Mochizuki2012,Onose2012,Schwarze2014} 
A natural question arises whether a similar mode spectrum exists for a single magnetic skyrmion within a spin-polarized background. A first numerical study of the magnon-skyrmion bound states has been recently carried out by Lin {\it et al.}.\cite{Lin2014} 
On the other hand, the scattering of magnons from a single skyrmion has been investigated with the help of micromagnetic simulations by Iwasaki {\it et al.}.\cite{Iwasaki2014} A characteristic skew scattering was found which was attributed to an emergent Lorentz force that is generated by the skyrmion topology. Moreover, the simulations revealed that the skyrmion experiences a magnon pressure pushing it towards the magnon source, which was explained in terms of momentum conservation.\cite{Iwasaki2014}

In the present work, we investigate the magnon fluctuations in the presence of a skyrmion texture analytically.
Starting from the non-linear sigma model description of a two-dimensional chiral magnetic system, we derive in Sec.~\ref{sec:Model} the magnon Hamiltonian by expanding around the skyrmionic saddle point solution. The  spectrum of the magnon-skyrmion bound states is obtained, and the magnon-skyrmion scattering cross section is analyzed in Sec.~\ref{sec:Fluctuations}. A theory for the magnon pressure, i.e., the momentum-transfer force exerted on the skyrmion by a magnon current, is presented in Sec.~\ref{sec:MagnonPressure}, which in particular explains previous numerical work. 
We conclude with a summary and discussion of the results in Sec.~\ref{sec:summary}.

\section{Theory for magnon-skyrmion scattering in a chiral magnet}
\label{sec:Model}
\subsection{Action of a two-dimensional chiral magnet}

Our starting point is the standard model of cubic chiral magnets.\cite{Bak80,Nakanishi80} We limit ourselves however to an  effective two-dimensional magnetic system thus restricting the spatial coordinate $\vect r$ to a plane. The orientation of the magnetisation $\vec M(\vect r) = M \hat n(\vect r)$ is parametrised by a unit vector $\hat n$ that is governed by the Euclidean action of the following non-linear sigma model
\begin{equation} \label{action}
S = \int_0^\beta d\tau \int d^2 \vect r\, \mathcal{L}\,\quad {\rm with}\quad
\mathcal{L} =  \mathcal{L}_{\rm dyn} + \mathcal{L}_{\rm stat}
\end{equation}
where $\beta = 1/(k_B T)$ is the inverse temperature. The Lagrangian consists of two parts. The static part reads
\begin{align}
\mathcal{L}_{\rm stat}
= &\frac{\varepsilon_{0}}{2} \Big[ (d_\alpha \hat n_i)^2 + 2 Q \epsilon_{i\alpha j} \hat n_i d_\alpha \hat n_j 
- 2 \kappa^2 \hat n \hat B \Big]
\label{Lstatic}
\end{align}
where here and in the following we use greek indices for two-dimensional real-space vectors, $\alpha = 1,2$ with e.g.~$d_1 = \frac{d}{d x}$, and latin indices for the magnetisation vector, $i,j,k = 1,2,3$. $\epsilon_{i k j}$ is the totally antisymmetric tensor with $\epsilon_{123} = 1$. The spin-orbit Dzyaloshinskii-Moriya interaction is proportional to $Q>0$ that we choose to be positive representing a right-handed chiral magnetic system. $\varepsilon_{0}$ is the energy scale associated with the stiffness and $\kappa>0$ measures the strength of the magnetic field. 
We will consider here only the situation when the magnetic field is orthogonal to the two-dimensional plane, $\hat B = \hat z$. In the following, cubic anisotropies\cite{Bak80,Nakanishi80} will be neglected because they are assumed to be weak and of only minor importance for the problem investigated here. We also neglect for simplicity dipolar interactions, which are known to give rise to quantitative and qualitative corrections though, e.g., for the magnetic resonances.\cite{Schwarze2014}

The dynamic part is given by the Berry phase\cite{AuerbachBook}
\begin{align}
\mathcal{L}_{\rm dyn} = - \frac{\hbar}{a^2} \vec A(\hat n) i d_\tau \hat n
\label{Ldyn}
\end{align}
where $a$ is a typical distance between the magnetic moments  and $\tau = i t$ is the imaginary time. The gauge field satisfies $\epsilon_{ijk}\partial \vec A_j/\partial \hat n_i = \hat n_k$, and the Euler-Lagrange equations of the action Eq.~(\ref{action}) reproduce the standard Landau-Lifshitz equations for the magnetisation. Note that we have chosen the sign of \eqref{Ldyn} such that the magnetization vector, $\hat n$, is antiparallel to the spin vector as it is the case for the magnetic moments of electrons.
We neglect throughout this work additional degrees of freedom that might give rise to dissipation usually represented by a phenomenological Gilbert damping. 

\begin{table}
\begin{tabular}{c|c}
\hline
stiffness energy & $\varepsilon_0$\\
Dzyaloshinskii-Moriya momentum &  $Q$ \\
spin density & $\hbar/a^2$ \\
skyrmion radius & $1/\kappa$\\
\hline
Dzyaloshinskii-Moriya energy &  $\varepsilon_{\rm DM} = \varepsilon_0 a^2 Q^2$ \\
magnon gap energy & $\varepsilon_{\rm gap} = \varepsilon_0 a^2 \kappa^2$\\
magnon mass & $M_{\rm mag} = \hbar^2/(2\varepsilon_0 a^2)$ \\
\hline
\end{tabular}
\caption{Parameters used throughout this work. The value of $\kappa$ can be tuned by the strength of the applied magnetic field.}
\label{tab:MagnonParameters}
\end{table}

\subsubsection{Length scales and parameters}
The model \eqref{action} possesses three in general different length scales: $a$, $1/\kappa$ and $1/Q$. We are interested here in the parameter regime where a single skyrmion is a topologically stable excitation of the magnetic system which is the case for intermediate values of the magnetic field, $\kappa \sim Q$. Whereas for small $\kappa \ll Q$ the magnetic system becomes unstable with respect to a proliferation of skyrmions, for very large values $\kappa \gg Q$ the energy density in the core of a skyrmion is very high so that amplitude fluctuations of the magnetisation become important eventually destroying the skyrmion.\cite{Ezawa2011} As we will see below, $1/\kappa$ can be identified with the skyrmion radius. Moreover, we consider the limit of small spin-orbit coupling implying $Q a \ll 1$ to be a small parameter. This implies, in particular, that a single skyrmion is composed of many individual spins. 

The magnon excitations of the field-polarised ground state are characterised by a gap and a mass that are given for later convenience in Table~\ref{tab:MagnonParameters}.

\subsubsection{Topological charge}

We now turn to the discussion of the conserved currents associated with the Lagrangian \eqref{action}. In the following, it will be convenient to use 2+1 dimensional space-time vectors, for which we use the indices $\mu,\nu,\lambda = 0,1,2$ with the time-derivative $d_0 = i d_\tau$. However, we still reserve the indices $\alpha,\beta = 1,2$ for spatial vectors only, see Table~\ref{tab:Indices}.

The order parameter $\hat n$ is an element of the two-dimensional sphere $S^2$. Its second homotopy group  is the group of integer numbers $\pi_2(S^2) = \mathbb{Z}$, and, as a result, the order parameter allows for topological {\it textures} in two spatial dimension. The associated 2+1 topological current vector reads 
\begin{align} \label{TopCurrent}
j^{\rm top}_\mu = \frac{1}{8\pi} \epsilon_{\mu \nu \lambda}\, \hat n (d_\nu \hat n \times d_\lambda \hat n),
\end{align}
where $\epsilon_{012} = 1$. In case the field configuration $\hat n$ is non-singular, i.e., in the absence of hedgehog defects in 2+1 space-time, this current is conserved
\begin{align} \label{ConservedTopCurrent}
d_\mu j^{\rm top}_\mu = 0,  
\end{align}
as $\hat n$ is a unit vector and $\hat n\, d_\nu \hat n = 0$. The topological charge density reads explicitly
\begin{align}
j^{\rm top}_0 &= \frac{1}{4\pi}  \hat n\, (d_1 \hat n \times d_2 \hat n).
\end{align}
The spatial integral over the charge density is quantised
\begin{align} \label{WindingNumber}
\int d^2 \vect r j^{\rm top}_0 \equiv W \in \mathds{Z},
\end{align}
and identifies the winding number $W$ of the texture. In this paper, we focus on the (baby-)skyrmion texture with $W= -1$.
For later convenience we note that the topological current can be expressed in terms of the spin-gauge field of Eq.~\eqref{Ldyn} as follows
\begin{align} \label{TopCurrentSpinGauge}
j^{\rm top}_\mu = 
\frac{1}{4\pi} \epsilon_{\mu \nu \lambda} (d_\nu \vec A) d_\lambda \hat n.
\end{align}

\begin{table}
\begin{tabular}{c|c}
\hline
spin vector indices & $i,j,k = 1,2,3$ \\
spatial indices & $\alpha, \beta, \gamma = 1,2$ \\
space-time indices & $\mu,\nu,\lambda = 0,1,2$\\
\hline
\end{tabular}
\caption{Indices used e.g.~in the discussion of conserved currents. For the time derivative we also use the notation $d_0 = d_t = i d_\tau$.}
\label{tab:Indices}
\end{table}

There are also important Noether currents of the Lagrangian $\mathcal{L}$ related to momentum and angular momentum conservation that are discussed in detail in appendix \ref{app:Noether}.

\subsection{Saddle point solution of the magnetic skyrmion}
\label{sec:saddlepointskyrmion}

For sufficiently large magnetic fields and, thus, large values of $\kappa$, the action \eqref{action} is minimised by the fully polarised state $\hat n = \hat z$. A large amplitude excitation of this state is the magnetic skyrmion texture with a winding number $W=-1$, see Eq.~\eqref{WindingNumber}. The skyrmion profile is parameterised by $\hat n_s^T = (\sin \theta \cos \varphi, \sin \theta \sin \varphi, \cos \theta)$ with 
\begin{align} \label{SkyrmionParameterization}
\varphi = \chi + \pi/2,\quad {\rm and}\quad \theta = \theta(\rho),
\end{align}
in terms of polar coordinates, $\rho$ and $\chi$, of the two-dimensional spatial distance vector $\delta \vect r^T = (\rho \cos \chi, \rho \sin \chi)$. 
The Euler-Lagrange equation deriving from Eq.~\eqref{Lstatic} yields the differential equation 
obeyed by the function $\theta(\rho)$,\cite{Bogdanov1989,Boganov1994}
\begin{align} \label{SkyrmionEquation}
\theta'' +\frac{\theta'}{\rho} - \frac{\sin\theta \cos\theta}{\rho^2} + \frac{2 Q \sin^2 \theta }{\rho} - \kappa^2 \sin \theta =0
\end{align}
with the boundary conditions $\theta(0)=\pi$ and $\lim_{\rho \to \infty} \theta(\rho) = 0$. Its solution possesses the asymptotics 
\begin{align} \label{SkyrmionAsymptotics}
\theta(\rho) \approx \left\{
\begin{array}{lll}
\pi - c_1 \kappa \rho & {\rm for} & \rho \to 0
\\
\frac{c_2}{\sqrt{\kappa \rho}} e^{-\kappa \rho} & {\rm for} &\rho \to \infty
\end{array}
\right.
\end{align}
with positive coefficients $c_1$ and $c_2$, that however depend on the ratio $\kappa/Q$. 
The exponential decay for large distances identifies $1/\kappa$ as the skyrmion radius. 

\begin{figure}
\centering
\includegraphics[width=0.9\columnwidth]{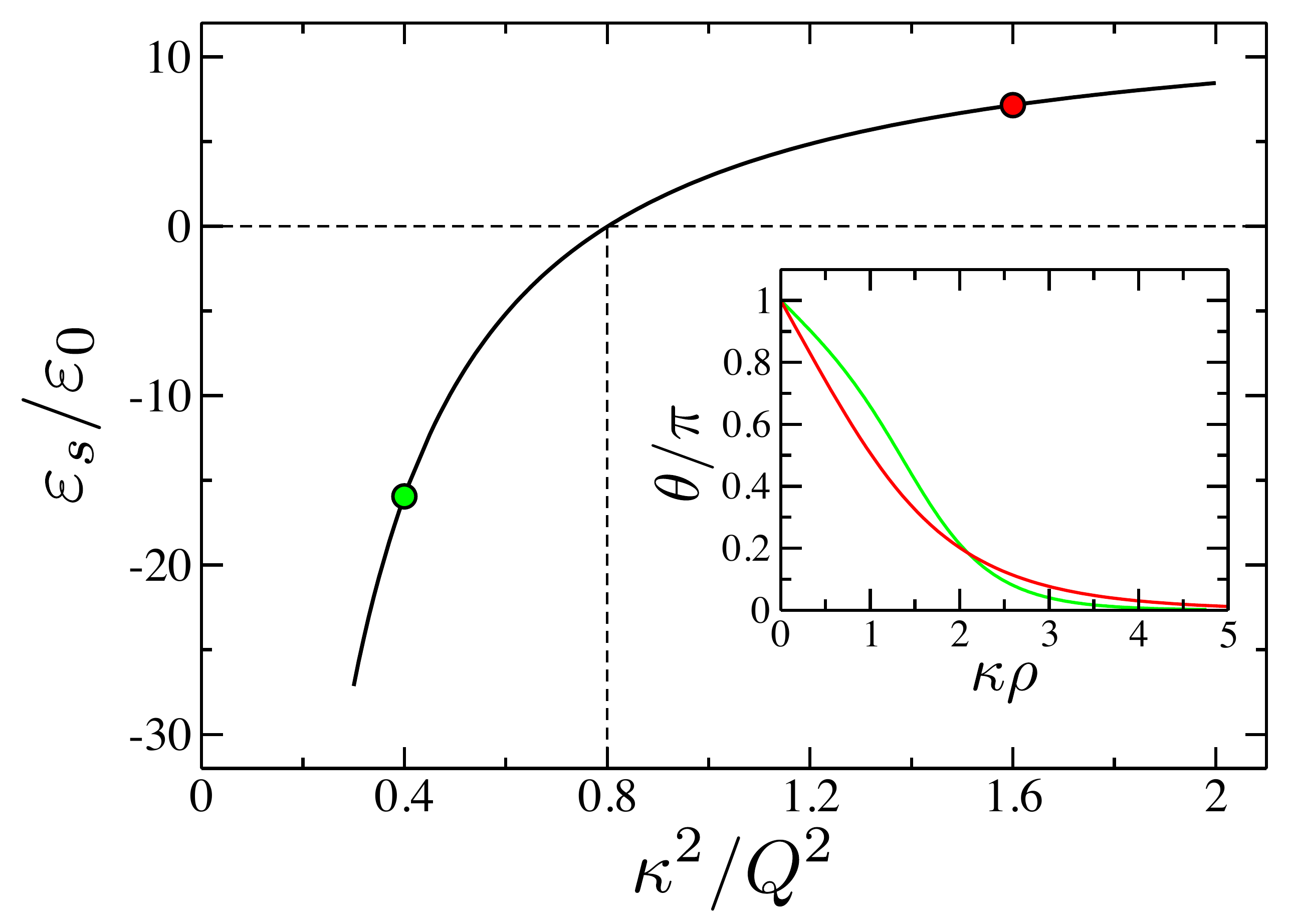}
\caption{Energy dependence of the single skyrmion solution as a function of the dimensionless parameter $\kappa^2/Q^2$. The energy of the skyrmion is positive for $\kappa^2 > \kappa_{\rm cr}^2 \approx 0.8 Q^2$. The negative skyrmion energy for smaller $\kappa$ signals an instability of the magnetic systems towards a proliferation of skyrmions. Inset shows two skyrmion profiles $\theta(\rho)$ for values of $\kappa^2/Q^2$ marked as dots in the main panel.}
\label{fig:skyrmionEnergy}
\end{figure}

The boundary value problem \eqref{SkyrmionEquation} is easily numerically solved with the help of the shooting method.\cite{morrison1962multiple} Here, the constant value $c_1$ of the short-distance asymptotics \eqref{SkyrmionAsymptotics} is varied until one obtains a monotonous function $\theta(\rho)$ with the required exponentially decaying behaviour at large distances. Examples of the numerically obtained profiles are shown in the inset of Fig.~\ref{fig:skyrmionEnergy}. The resulting skyrmion texture is illustrated in Fig.~\ref{fig:skyrmionTexture}.

Integrating the static skyrmion solution one obtains for its saddle-point action $S^{(0)}_s =  \beta (\varepsilon_s + \varepsilon_{\rm FP})$ where $\varepsilon_{\rm FP} = - \varepsilon_0 \kappa^2 \mathcal{V}$ is the energy of the field-polarized state, $\hat n_{\rm FP} \equiv \hat z$, with the volume $\mathcal{V}$ and  $\varepsilon_s = \varepsilon_{0} \mathcal{E}(\kappa^2/Q^2)$ identifies the energy of the skyrmion. The dimensionless function $\mathcal{E}$ is shown in Fig.~\ref{fig:skyrmionEnergy}. The energy of the skyrmion is positive as long as $\kappa > \kappa_{\rm cr}$ where
\begin{align} \label{GlobalStability}
\kappa_{\rm cr}^2 \approx 0.8 Q^2. 
\end{align}
For values smaller than the critical value $\kappa_{\rm cr}$ the magnetic system becomes unstable towards a proliferation of skyrmions and, eventually, the formation of a skyrmion lattice. In the following, we concentrate on the regime with positive skyrmion energy, $\kappa > \kappa_{\rm cr}$, where the skyrmion is a large amplitude excitation of the field-polarised phase with a positive excitation energy.

\subsection{Effective theory for magnon excitations in the presence of a  skyrmion}

We now consider spin-wave fluctuations, i.e., the magnon modes in the presence of a single skyrmion in a chiral magnet. A similar analysis for magnetic solitons and vortices in ferromagnets has been carried out previously.\cite{Kosevich1990,Ivanov1995,Ivanov1998,Sheka2001} For the parameterisation of the magnons we introduce the local orthogonal frame $\hat e_i(\vect r) \hat e_j(\vect r) = \delta_{ij}$ with $i,j = 1,2,3$ and $\hat e_1(\vect r) \times \hat e_2(\vect r) = \hat e_3(\vect r)$ where $\hat e_3(\vect r) = \hat n_s(\vect r)$ follows the static skyrmion profile. We will use the following representation in terms of polar and azimuthal angle, $\theta$ and $\varphi$, respectively, introduced above
\begin{align} 
\nn
\hat e_{1}^T(\vect r) &= (-\sin\varphi,  \cos\varphi, 0)
\\
\label{dreibein}
\hat e_{2}^T(\vect r) &= (-\cos\theta \cos\varphi, - \cos\theta \sin\varphi, \sin\theta)
\\\nn
\hat e_{3}^T(\vect r) &= (\sin \theta \cos \varphi, \sin \theta \sin \varphi, \cos \theta),
\end{align}
where $\varphi = \chi + \pi/2$ and $\theta = \theta(\rho)$. Furthermore, it is convenient to introduce the chiral vectors
\begin{align}
\hat e_{\pm} = \frac{1}{\sqrt{2}} (\hat e_{1} \pm i \hat e_{2})
\end{align}
that have the property $\hat e_{+}\hat e_{+} = \hat e_{-} \hat e_{-} = 0$ and $\hat e_{+}\hat e_{-} =  \hat e_{-}\hat e_{+} = 1$.

Due to translation invariance, the energy of the skyrmion is independent of its position giving rise to two zero modes
associated with translations of the skyrmion.\cite{RemarkUmklapp} In order to treat these zero modes we introduce two time-dependent collective coordinates $\vect R^T(\tau) = (R_x(\tau),R_y(\tau))$. The remaining massive fluctuation modes are represented by the dimensionless complex field $\psi(\vect r, \tau)$. We use the parameterization
\begin{align}
\hat n(\vect r,\tau) &= \hat e_3(\vect r - \vect R(\tau)) \sqrt{1- 2 |\psi(\vect r - \vect R(\tau),\tau)|^2} \nonumber
\\
&+ \hat e_+(\vect r - \vect R(\tau)) \psi(\vect r - \vect R(\tau),\tau) 
\\\nn
&+ \hat e_-(\vect r - \vect R(\tau)) \psi^*(\vect r - \vect R(\tau),\tau).
\end{align}
The orthogonal frame $\hat e_i $ depends on the radius $\rho$ and the angle $\chi$ that are here associated with the distance vector $\delta \vect r^T = (\vect r - \vect R)^T = (\rho \cos \chi, \rho \sin \chi)$.

It is clear that this parameterization is invariant under the local transformation
\begin{align} \label{GaugeTrafo}
&\hat e_+ \to \hat e_+ e^{-i \lambda},\qquad \psi \to \psi e^{i \lambda}
\\
&\hat e_- \to \hat e_- e^{i \lambda},\qquad \psi^* \to \psi^* e^{-i \lambda}
\end{align}
with  $\lambda = \lambda(\vect r - \vect R(\tau),\tau)$. In the limit of large distances from the skyrmion, $|\vect r - \vect R(\tau)| \to \infty$, the parameterisation assumes the form
\begin{align}
\hat n(\vect r,\tau) &\approx \hat z \sqrt{1- 2 |\psi(\vect r - \vect R(\tau),\tau)|^2} \nonumber
\\
&+ \frac{1}{\sqrt{2}} (\hat x + i \hat y) \Big[- i e^{- i \varphi} \psi(\vect r - \vect R(\tau),\tau)\Big]
\\\nn
&+ \frac{1}{\sqrt{2}} (\hat x - i \hat y) \Big[i e^{i \varphi} \psi^*(\vect r - \vect R(\tau),\tau)\Big]
\end{align}
The last two lines allow to identify the magnon wave function with respect to an orthogonal frame that becomes the laboratory frame, $\hat x$, $\hat y$ and $\hat z$, at large distances,
\begin{align} \label{LabFrame}
\psi_{\rm LAB}(\vect r - \vect R(\tau),\tau) = - i e^{- i \varphi} \psi(\vect r - \vect R(\tau),\tau).
\end{align}
This will become important later for the discussion of magnon scattering off the skyrmion.

Expanding the action \eqref{action} in the massive field $\psi$ allows us to study their properties and their influence on the skyrmion. 

\subsubsection{Zeroth order in the massive fluctuation field $\psi$}
\label{subsubsec:ZerothOrderInPsi}

In zeroth order in the fluctuation field $\psi$ the Lagrangian is given by $\int d^2 \vect r \mathcal{L}^{(0)} = \varepsilon_{\rm FP} + \varepsilon_s + L^{(0)}$ with 
\begin{align} \label{Lagrangian0}
L^{(0)} = 
-\mathcal{A}(\vect R)  i d_\tau \vect R.
\end{align}
This term originates from the expansion of the dynamical part of the action (\ref{Ldyn}) and has the form of a massless particle
with coordinate $\vect R$ in the presence of a gauge field $\mathcal{A}$ given by 
\begin{align} \label{EffGaugeField}
\mathcal{A}_\alpha(\vect R) = 
-\frac{\hbar}{a^2} \int d^2 \vect r \vec A(\hat e_3(\vect r-\vect R)) \partial_\alpha \hat e_3(\vect r-\vect R).
\end{align}
The electric field associated with this vector potential vanishes as $\partial_\tau \mathcal{A}_\alpha(\vect R) = 0$. The effective magnetic field however is finite and determined by the skyrmion number
\begin{align}
\mathcal{B}_{\rm eff} &\equiv \epsilon_{z \alpha\beta}\frac{\partial}{\partial \vect R_\alpha}\mathcal{A}_\beta(\vect R) 
\\ \nonumber
&= \frac{\hbar}{a^2}  \int d^2 \vect r\, \hat n_s (\partial_1 \hat n_s \times \partial_2 \hat n_s)
= -\frac{4\pi\hbar}{a^2}.
\end{align}

Note that in zeroth order in the magnon field $\psi$, the gauge field \eqref{EffGaugeField} just coincides 
with the total canonical momentum\cite{Haldane1986}
\begin{align} \label{SkyrmionMomentum}
-\frac{\partial L^{(0)}}{\partial (i d_\tau \vect R_\alpha)} = \mathcal{A}_\alpha(\vect R) = \int d^2 \vect r\, T_{0\alpha}\Big|_{\psi,\psi^*=0},
\end{align}
where the energy-momentum tensor, $T_{\mu\nu}$, is defined in Eq.~\eqref{EMTensor}.
The classical equations of motion deriving from Lagrangian \eqref{Lagrangian0} have the form of a massless particle in a magnetic field, 
\begin{align} \label{ClassicalEoM}
{\vect  G}  \times d_t \vect R = 0,
\end{align}
with $\vect  G = \mathcal{B}_{\rm eff} \hat z$. They are to be identified with the well-known Thiele equations for magnetic textures in the absence of a Gilbert damping, and $\vect  G$ is the so-called  gyrocoupling vector.\cite{Thiele1973}
At zeroth order in $\psi$, these equations of motion also coincide with Eq.~\eqref{IntTopCurrent} following from momentum conservation implying that the integrated topological current vanishes. 

\subsubsection{Second order in the massive fluctuations $\psi$}

We now turn to the magnon spectrum and its interaction with the skyrmion. The scattering of magnons from the skyrmion does not conserve the magnon number, which is reflected in the presence of local, anomalous quadratic terms $\psi \psi$ and $\psi^* \psi^*$ in the Hamiltonian.  It is therefore convenient to use the spinor notation
\begin{align}
\vec \psi = \left(
\begin{array}{c}
\psi \\ \psi^*
\end{array}
\right),\qquad
\vec \psi^\dagger = (\psi^*, \psi).
\end{align}
Expanding the Lagrangian density in second order in the fluctuation field $\psi$ one finds after some algebra
\begin{align} \label{Lagrangian2}
\mathcal{L}^{(2)} =
\frac{1}{2 a^2}  \vec \psi^\dagger \tau^z \hbar \partial_\tau \vec \psi
+ \frac{1}{2 a^2} \vec \psi^\dagger H \vec \psi + \mathcal{L}^{(2)}_{\rm int}
\end{align}
where $\tau^z$ is a Pauli matrix and we used that the field $\psi = \psi(\vect r - \vect R(\tau),\tau)$ possesses an explicit and implicit time-dependence, $d_\tau \psi = \partial_\tau \psi - (\partial_\gamma \psi) d_\tau \vect R_\gamma$.
The bosonic Bogoliubov-deGennes Hamiltonian reads
\begin{align} 
\label{Hamiltonian1}
&H = \\\nn
&\varepsilon_{0} a^2
\Big[- \mathds{1} \nabla^2+ 2 \tau^z  \left(\frac{\cos \theta}{\rho^2} 
- Q \frac{\sin \theta}{\rho} \right) i \partial_\chi
+ \mathds{1} V_0 + \tau^x V_x
\Big].
\end{align}
The potentials are given by
\begin{align}
V_0 &= 
\frac{1+3\cos(2\theta)}{4 \rho^2} - \frac{3 Q \sin(2 \theta)}{2\rho} + \kappa^2 \cos \theta
- Q \theta' - \frac{1}{2}\theta'^2
\nn\\
V_x &= \frac{\sin^2(\theta)}{2 \rho^2} + \frac{Q \sin(2 \theta)}{2\rho} - Q \theta' - \frac{1}{2}\theta'^2,
\end{align}
and only depend on the distance $\rho$ to the skyrmion center as $\theta = \theta(\rho)$.
Here, $\nabla^2$ is the two-dimensional Laplace operator which reads in polar coordinates $\nabla^2 = \partial_\rho^2 + (1/\rho) \partial_\rho + \partial_\chi^2/\rho^2$. For later reference, we note that the derivative in polar coordinates is given by $\partial_\alpha = \hat \rho_\alpha\, \partial_\rho + \hat \chi_\alpha \frac{\partial_\chi}{\rho}$ with the unit vectors 
\begin{align}
\hat \rho^T = (\cos \chi, \sin \chi),\quad  \hat \chi^T = (- \sin \chi, \cos \chi).
\end{align}
By completing the square, the Hamiltonian can be also written in the form 
\begin{align} 
\label{Hamiltonian2}
H &= \varepsilon_{0} a^2
\Big[
(-i \nabla - \tau^z \vec a)^2  
+ \mathds{1} (V_0 - \vec a^2) + \tau^x V_x
\Big],
\end{align}
with the gauge field
\begin{align} 
\label{GaugeField}
\vec a  &=  \left(\frac{\cos \theta}{\rho} - Q \sin \theta \right) \hat \chi.
\end{align}
Note that $\vec a$ assumes here the Coulomb gauge $\nabla \vec a = 0$.

The interaction between the magnon field $\vec \psi$ and the zero mode is given by
\begin{align} \label{MagnonVelocityInteraction}
\mathcal{L}^{(2)}_{\rm int} &=
- \frac{\hbar}{2 a^2} \vec \psi^\dagger \Gamma^\gamma \vec \psi\, i d_\tau \vect R_\gamma.
\end{align}
The interaction vertex reads
\begin{align} \label{InteractionVertex}
\Gamma^\gamma &= 
- \tau^z i \partial_\gamma - \mathds{1} \frac{\cos \theta}{\rho} \hat \chi_\gamma.
\end{align}
Whereas the first term just combines  with the dynamic part of the Lagrangian \eqref{Lagrangian2} to a total time derivative, the second term derives from a spin connection attributed to the local orthogonal frame \eqref{dreibein}.

If the skyrmion velocity vanishes, $i d_\tau \vect R = 0$, the Lagrangian reduces to the first two terms 
in Eq.~\eqref{Lagrangian2}. These terms constitute a Bogoliubov-deGennes scattering problem and determine the spectrum of the magnons in the presence of a skyrmion with $i d_\tau \vect R = 0$. It is important to note that these terms and thus the spectrum do not depend on the position of the skyrmion $\vect R$ itself as the collective coordinate $\vect R$ can be eliminated by a change of the integration variable $\vect r - \vect R \to \vect r$ (Ref.~\onlinecite{RajaramanBook}).


\section{Magnon spectrum and skyrmion scattering cross section} 
\label{sec:Fluctuations}

In the following we investigate the properties of the magnons by analysing the eigenvalues and eigenstates of the Hamiltonian \eqref{Hamiltonian1}. It requires the solution of a Bogoliubov-deGennes scattering problem, which then allows us to determine the magnon-skyrmion bound states and to address the magnon scattering cross section for the case of a vanishing skyrmion velocity, $i d_\tau \vect R = 0$. 

\subsection{Bogoliubov-deGennes scattering problem}

The magnons are obtained as eigenstates of the eigenvalue problem
\begin{align}
H \vec \psi = \varepsilon \tau^z \vec \psi,
\end{align}
with the Hamiltonian $H$ given in Eq.~\eqref{Hamiltonian1}. 
The Hamiltonian possesses the following particle-hole symmetry 
\begin{align}
\tau^x K H \tau^x K = H
\end{align}
where $K$ is complex conjugation, which originates in the fact that the magnetisation is a real quantity. 
As a consequence, the spectrum of $H$ is characterized by pairs $\pm \varepsilon$ of eigenvalues. In particular, if $\vec \psi$ is an eigenvector with eigenvalue $\varepsilon$ then $\tau^x K \vec \psi$
is an eigenvector with eigenvalue $-\varepsilon$.

Setting $\vec \psi = e^{i m \chi} \vec \eta_m(\rho)$ with the angular momentum quantum number $m$ and using $H = H(- i \partial_\chi)$ the eigenvalue equation reduces to
\begin{align} \label{EigenvalueEq}
H(m) \vec \eta_m = \varepsilon \tau^z \vec \eta_m.
\end{align}
%
The spectrum of $H$ contains discrete bound states labeled by the quantum number $n$ and eigenfunctions $\vec \eta_{m,n}$ as well as scattering states labeled by the energy $\varepsilon$ and eigenfunctions $\vec \eta_{m,\varepsilon}$. These eigenfunctions will be normalized such that 
\begin{align} \label{NormalizationBound}
\int_0^\infty d\rho \rho\, \vec \eta^\dagger_{m,n} \tau^z  \vec \eta_{m,n'} &= \delta_{n,n'},
\\\label{NormalizationScattering}
\int_0^\infty d\rho \rho\, \vec \eta^{\dagger}_{m,\varepsilon} \tau^z  \vec \eta_{m,\varepsilon'} 
&= \delta(\varepsilon- \varepsilon').
\end{align}
Stability of the theory \eqref{Lagrangian2} then requires the energy to be positive $\varepsilon\geq0$. 
%
%
The eigenvectors with negative eigenenergies, $-\varepsilon \leq 0$, are then given by $\tau^x K e^{i m \chi} \vec \eta_m = e^{-i m \chi} \vec \zeta_{-m}$ where $\vec \zeta_{-m} = \tau^x \vec \eta^*_m$ and 
\begin{align}
H(-m) \vec \zeta_{-m} = -\varepsilon \tau^z \vec \zeta_{-m}.
\end{align}

\subsubsection{Definition of the skyrmion scattering problem}

\begin{figure}
\centering
\includegraphics[width=0.8\columnwidth]{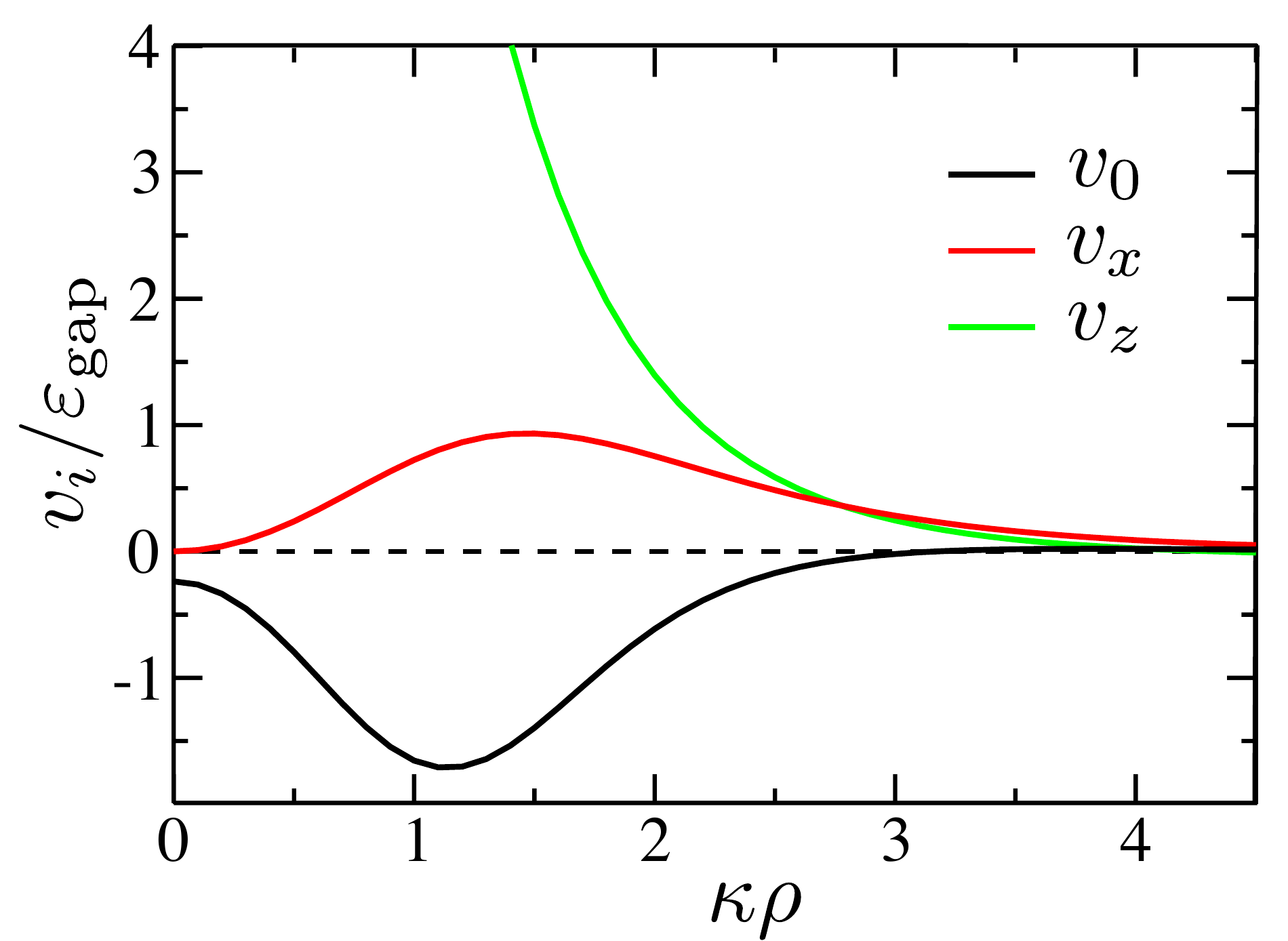}
\caption{The potentials $v_0$, $v_x$ and $v_z$ of Eqs.~\eqref{ScatteringPots} in units of $\varepsilon_{\rm gap}$
plotted as functions of the dimensionless parameter $\kappa\rho$ for a value of $\kappa=Q$. All three potentials vanish exponentially for $\kappa \rho \gg 1$.}
\label{fig:potentials}
\end{figure}

The skyrmion profile decays exponentially and its scattering potential is thus well localised.
We recast the Hamiltonian in terms of a scattering problem $H(m) = H_{0m} + \mathcal{V}_m$. Here $H_{0m}$ describes the magnons in the absence of the skyrmion
\begin{align}
H_{0m} = &\varepsilon_{0} a^2
\Big[\mathds{1} \left(-\partial_\rho^2 - \frac{\partial_\rho}{\rho} + \frac{m^2+1}{\rho^2} + \kappa^2 \right) 
- \tau^z \frac{2 m}{\rho^2} 
\Big],
\label{H0m}
\end{align}
and the skyrmion matrix scattering potential $\mathcal{V}_m = \mathcal{V}_m(\rho)$ is given by
\begin{align}
&\mathcal{V}_m(\rho)=
\varepsilon_{0} a^2
\Big[v_z(\rho) \tau^z  + v_0(\rho) \mathds{1} + v_x(\rho) \tau^x\Big] 
\label{Vm}
\end{align}
with
\begin{align} \label{ScatteringPots}
v_z(\rho) &=  - 2 m \left(\frac{\cos \theta-1}{\rho^2} - \frac{Q \sin \theta}{\rho} \right), 
\nn\\
v_0(\rho) &= \frac{3 (\cos(2\theta)-1)}{4 \rho^2} - \frac{3Q \sin(2\theta)}{2 \rho} + \kappa^2 (\cos \theta-1)
\nn\\&\qquad- Q \theta' - \frac{\theta'^2}{2},
\\\nn
v_x(\rho) &=\frac{\sin^2(\theta)}{2 \rho^2} + \frac{Q \sin(2\theta)}{2 \rho} - Q \theta' - \frac{\theta'^2}{2}. 
\end{align}
The potential $\mathcal{V}_m$ vanishes exponentially for $\kappa \rho \gg 1$ as illustrated in Fig.~\ref{fig:potentials}. The anomalous potential $v_x$ that couples the two components of the wave function $\vec \eta$ vanishes quadratically for $\kappa \rho \to 0$ and exponentially for $\kappa \rho \to \infty$.

Solutions of the free problem $H_{0m} \vec \eta^{(0)}_m = \varepsilon \tau^z \vec \eta^{(0)}_m$ only exist for energies
\begin{align} \label{MagnonDispersion}
\varepsilon = \varepsilon_0 a^2 (\kappa^2 + k^2) \equiv \varepsilon_{\rm gap} + \frac{\hbar^2 k^2}{2 M_{\rm mag}}
\end{align}
with the radial momentum $k \geq 0$. This identifies the magnon gap $\varepsilon_{\rm gap} = \varepsilon_0 a^2 \kappa^2$ and the magnon mass $\hbar^2/(2 M_{\rm mag}) = \varepsilon_0 a^2$, see Table~\ref{tab:MagnonParameters}. The eigenfunctions are given by
\begin{align}
\vec \eta^{(0)}_{m,\varepsilon} =
\left( \begin{array}{c} 1 \\ 0 \end{array} \right)
\frac{1}{\sqrt{2\varepsilon_0 a^2}} J_{m-1}(k \rho) .
\end{align}
where $J_\nu$ are Bessel functions of the first kind. They are normalized such that 
\begin{align}
&\int_0^\infty d\rho \rho\, \vec \eta^{(0) \dagger}_{m,\varepsilon} \tau^z  \vec \eta^{(0)}_{m,\varepsilon'} 
\\&
= \frac{1}{2\varepsilon_0 a^2}
\int_0^\infty d\rho \rho  J_{m-1}(k \rho)J_{m-1}(k' \rho) 
= \delta(\varepsilon- \varepsilon') \nonumber
\end{align}
where we used the  completeness relation of the Bessel functions 
\begin{align}
\delta(k-k') &= k \int_0^\infty d\rho \rho  J_{\nu}(k \rho)J_{\nu}(k' \rho). 
\end{align}

\subsubsection{Asymptotics of eigenfunctions and scattering phase shifts}

For small distances $\rho \kappa \ll 1$, the Hamiltonian reduces to 
\begin{align}
H(m) \approx \varepsilon_{0} a^2
\Big[\mathds{1} \left(- \partial_\rho^2 - \frac{\partial_\rho}{\rho} + \frac{m^2+1}{\rho^2}\right)
+ 2 \tau^z \frac{m}{\rho^2} 
\Big]
\label{approxH}
\end{align}
where we have omitted all terms of order $\mathcal{O}(1/(\kappa \rho))$. Note that the presence of the skyrmion inverts the sign of the linear $m$-term for $\kappa \rho \ll 1$ as compared to the free Hamiltonian \eqref{H0m}. From Eq.~(\ref{approxH}) follows the asymptotics of the eigenfunction at small distances
\begin{align} \label{AsClose}
\vec \eta_m \approx \left(
\begin{array}{c} c_3 (\kappa \rho)^{|m+1|} \\c_4 (\kappa \rho)^{|m-1|} 
\end{array} \right)
\quad {\rm for} \quad \kappa \rho \ll 1,
\end{align}
with constant coefficients $c_3$ and $c_4$.

The large distance asymptotics is governed by the free Hamiltonian \eqref{H0m} and depends on the energy $\varepsilon$. For energies below the magnon gap $\varepsilon < \varepsilon_{\rm gap}$ the wave function decays exponentially in $\kappa \rho$. For energies $\varepsilon = \varepsilon_{\rm gap} + \varepsilon_0 a^2 k^2$ with $k \geq 0$, the large distance asymptotics, $\kappa \rho \gg 1$, is given by
\begin{align} \label{AsLarge}
&\vec \eta_m \approx 
\\\nn&
\left( \begin{array}{c} 1 \\ 0 \end{array} \right)
\frac{1}{\sqrt{2\varepsilon_0 a^2}} 
\left(\cos ( \delta_m ) J_{m-1}(k \rho) - \sin ( \delta_m ) Y_{m-1}(k \rho)\right),
\end{align}
where $Y_\nu$ are the Bessel functions of the second kind, and we introduced the phase shift $\delta_m$. The second component is exponentially small and has been set to zero.

\subsection{Magnon-skyrmion bound states}

In the following we discuss the magnon bound states to be found within the energy range $0 \leq \varepsilon < \varepsilon_{\rm gap}$. 

\subsubsection{Cross-check: zero modes}

Before turning to a discussion of the bound states, however, we first perform an important cross-check. Although we are only interested here in the massive modes with finite energy, the spectrum of the Hamiltonian must also possess the zero modes corresponding to infinitesimal translations of the skyrmion, i.e., $\hat e_3(\vect r - \vect R(\tau)) \approx  \hat e_3(\vect r) - \partial_\gamma \hat e_3(\vect r)\vect R_\gamma(\tau)$. With the help of Eqs.~\eqref{dreibein} we find
\begin{align}
\partial_\gamma \hat e_3 = - \theta' \hat \rho_\gamma \hat e_2 + \frac{\sin \theta}{\rho} \hat \chi_\gamma \hat e_1.
\end{align}
This allows to identify a zero mode with angular momentum $m = -1$,
\begin{align}
\vec \eta^{\rm \,zm}_{-1} =\frac{1}{\sqrt{8}} \Big(\begin{array}{c} 
\frac{\sin \theta}{\rho} - \theta' \\
\frac{\sin \theta}{\rho} + \theta'
\end{array}\Big),
\end{align}
normalized according to Eq.~\eqref{NormalizationBound}, and the corresponding partner $\vec \zeta^{\rm \,zm}_{1} = \tau^x (\vec \eta^{\rm \,zm}_{-1})^*$. Using the differential equation \eqref{SkyrmionEquation} obeyed by $\theta(\rho)$ one can check explicitly that indeed $H(-1) \vec \eta^{\rm \,zm}_{-1} = H(1) \vec \zeta^{\rm \,zm}_{1} = 0$.
Translations are then described by the wavefunction $\vec\Psi^{\rm zm} = a\, \vec \eta^{\rm \,zm}_{-1} e^{-i \chi} + a^* \vec \zeta^{\rm \,zm}_{1} e^{i \chi}$ where the real and imaginary part of the coefficient $a \in \mathds{C}$ parametrize translations in $y$- and $x$-direction, respectively. Although there are two translational zero modes, it should be kept in mind that they are represented by a single zero mode in the spectrum obtained from the eigenvalue equation \eqref{EigenvalueEq} for the $\vec \eta_m$ wavefunctions.

\subsubsection{Massive magnon-skyrmion bound states}

The massive bound states with finite energy are determined by solving the eigenvalue equation \eqref{EigenvalueEq} numerically using again the shooting method.\cite{morrison1962multiple} This is done by first choosing a value for the ratio $c_3/c_4$ in Eq.~\eqref{AsClose} as well as the energy $\varepsilon$ and afterwards checking whether the numerically solution of the differential equation allows for a bound state that is exponentially decaying at large distances. This is iteratively repeated until a bound state is found. The absolute value for $c_3$ is afterwards fixed by the normalization condition \eqref{NormalizationBound}.
%
%

\begin{figure}
\centering
\includegraphics[width=0.9\columnwidth]{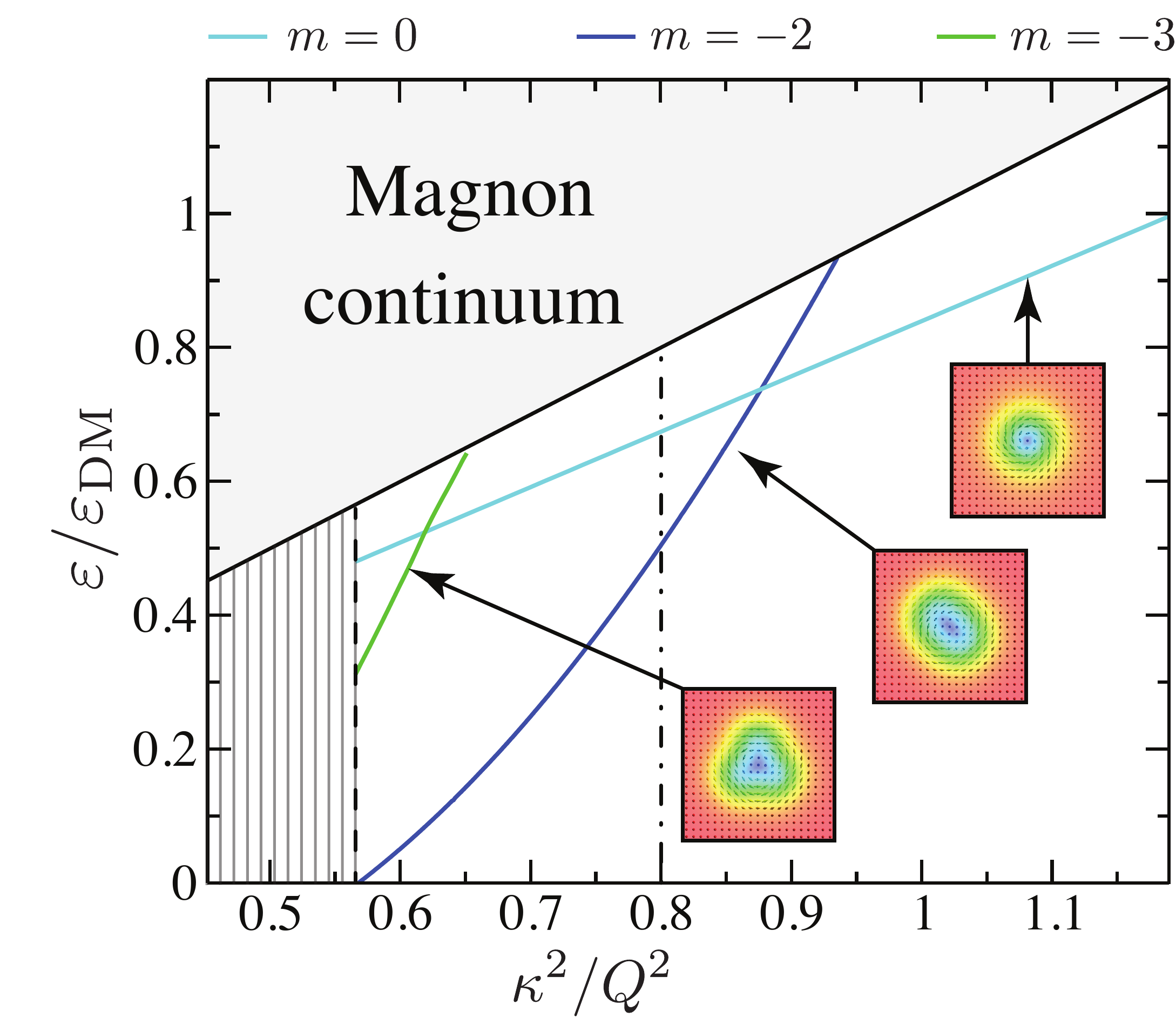}
\caption{Energy spectrum of magnons in the presence of a single skyrmion excitation of the field-polarised ground state. The latter becomes thermodynamically unstable for $\kappa^2 < \kappa_{\rm cr}^2 \approx 0.8 Q^2$ (dashed-dotted vertical line), see Eq.~\eqref{GlobalStability}. 
In addition to the continuous magnon spectrum for $\varepsilon > \varepsilon_{\rm gap} = \varepsilon_{\rm DM} \kappa^2/Q^2$, one obtains in-gap bound states for smaller energies. 
%
The dashed vertical line signals the local bimeron instability identified by the vanishing of the quadrupolar eigenfrequency, see text. 
The images show snapshots of the corresponding excitation modes, see also Fig.~\ref{fig:BoundStatesVis}.
 }
 \label{fig:spectrum}
\end{figure}

\begin{figure}
 \centering
\includegraphics[width=0.9\columnwidth]{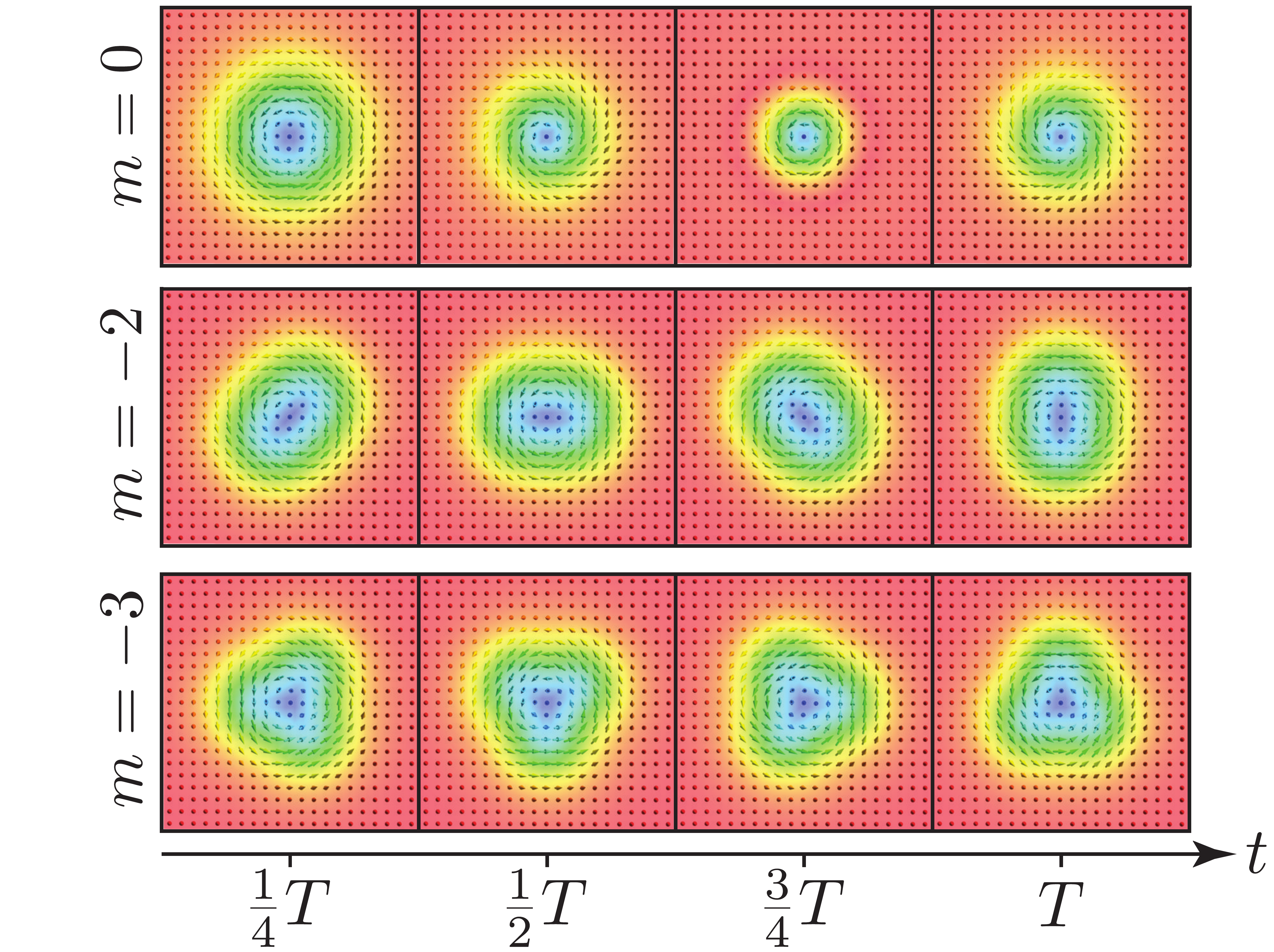}
\caption{Time-dependence of the skyrmion profile for the bound magnon modes with oscillation period $T = \hbar/\varepsilon$ where $\varepsilon$ is the corresponding eigenenergy. The colour code reflects the $z$-component of the local magnetisation. Corresponding movies can be found in Ref.~\onlinecite{Movies}.
}
 \label{fig:BoundStatesVis}
\end{figure}

The resulting energy spectrum is shown in Fig.~\ref{fig:spectrum}. In the regime of a stable field-polarized state $\kappa^2 > \kappa_{\rm cr}^2 \approx 0.8 Q^2$, see Fig.~\ref{fig:skyrmionEnergy}, we find two bound states in addition to the zero mode that is not shown. There exists a bound breathing mode with $m=0$ for all $\kappa$, and an additional bound quadrupolar mode with $m=-2$ appears below $\kappa^2 \lesssim 0.93 Q^2$. The eigenfrequency of this $m=-2$ mode decreases with decreasing $\kappa$ and eventually turns negative for values of $\kappa^2$ smaller than 
%
\begin{align}
\kappa_{\rm bimeron}^2 \approx 0.56 Q^2.
\end{align}
This negative energy eigenvalue corresponds to a local instability of the theory \eqref{Lagrangian2}, and it translates to an instability of the single skyrmion with respect to a static quadrupolar deformation. Such a deformed skyrmion can also be identified as a bimeron, and this bimeron instability was previously pointed out by Ezawa.\cite{Ezawa2011-2}
An additional $m=-3$ mode materializes in the intermediate, thermodynamically metastable regime, $\kappa^2_{\rm bimeron} < \kappa^2 < \kappa_{\rm cr}^2$, and bound states with higher $|m|$ 
do not exist in the locally stable regime $\kappa^2 > \kappa^2_{\rm bimeron}$.
Fig.~\ref{fig:BoundStatesVis} illustrates the space-time dependence of the relevant bound magnon modes, for movies see Ref.~\onlinecite{Movies}. 
%
%
%
All wave functions $\vec \eta_m(\rho)$ for the bound states we have found numerically do not possess any nodes, i.e., they do not have any zeros for some finite value of $\rho$. We were not able to find bound states with a single or more nodes.

The magnon spectrum of Fig.~\ref{fig:spectrum} agrees nicely with recent results of finite-size diagonalization of the Landau-Lifshitz-Gilbert equation by Lin {\it et al.}.\cite{Lin2014} The only qualitative difference seems to be a hybridisation of the $m=0$ and $m=-2$ mode close to their crossing at $\kappa^2 \approx 0.875 Q^2$ for the finite size system investigated in Ref.~\onlinecite{Lin2014}.

\subsection{Magnon scattering states}

Magnon scattering states are obtained for energies $\varepsilon \geq \varepsilon_{\rm gap}$.
The corresponding wave functions are also obtained numerically with the help of the shooting method. For a fixed energy $\varepsilon$ one first finds a value for the ratio $c_3/c_4$ of the small distance asymptotics, Eq.~\eqref{AsClose}, so that the numerical solution for the wave function $\vec \eta_m$ possesses a second component decaying exponentially with distance in agreement with Eq.~\eqref{AsLarge}. In a second step, the absolute value of, e.g., $c_3$ is then fixed so that the large distance asymptotics of the first component is consistent with Eq.~\eqref{AsLarge}, which ensures that the scattering wave functions are normalised according to Eq.~\eqref{NormalizationScattering}.

\begin{figure}
 \centering
\includegraphics[width=0.9\columnwidth]{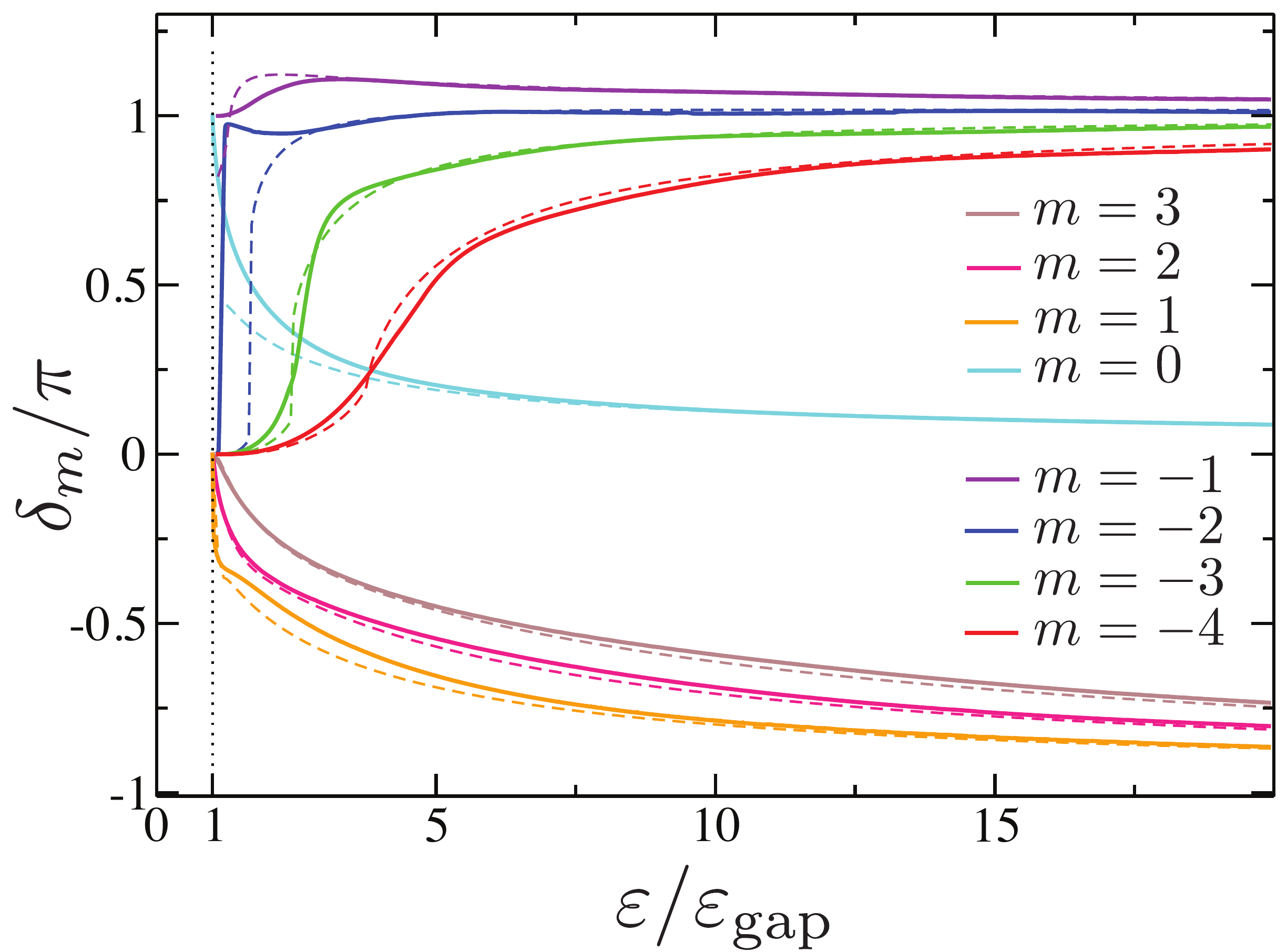}
\caption{Phase shifts of the scattering states for various angular momenta $m$ for $\kappa = Q$; numerically exact values are shown as solid lines, and the dashed lines are obtained within the WKB approximation. 
For large angular momenta or high energies the WKB approximation provides satisfying results. 
} \label{fig:phaseshift}
\end{figure}

The energy dependence of the numerically obtained phase shifts, $\delta_m$, are shown for the lowest angular momenta as solid lines in Fig.~\ref{fig:phaseshift} for $\kappa = Q$. 

\subsubsection{WKB approximation for the phase shift}

\begin{figure}
\centering
\includegraphics[width=0.48\columnwidth]{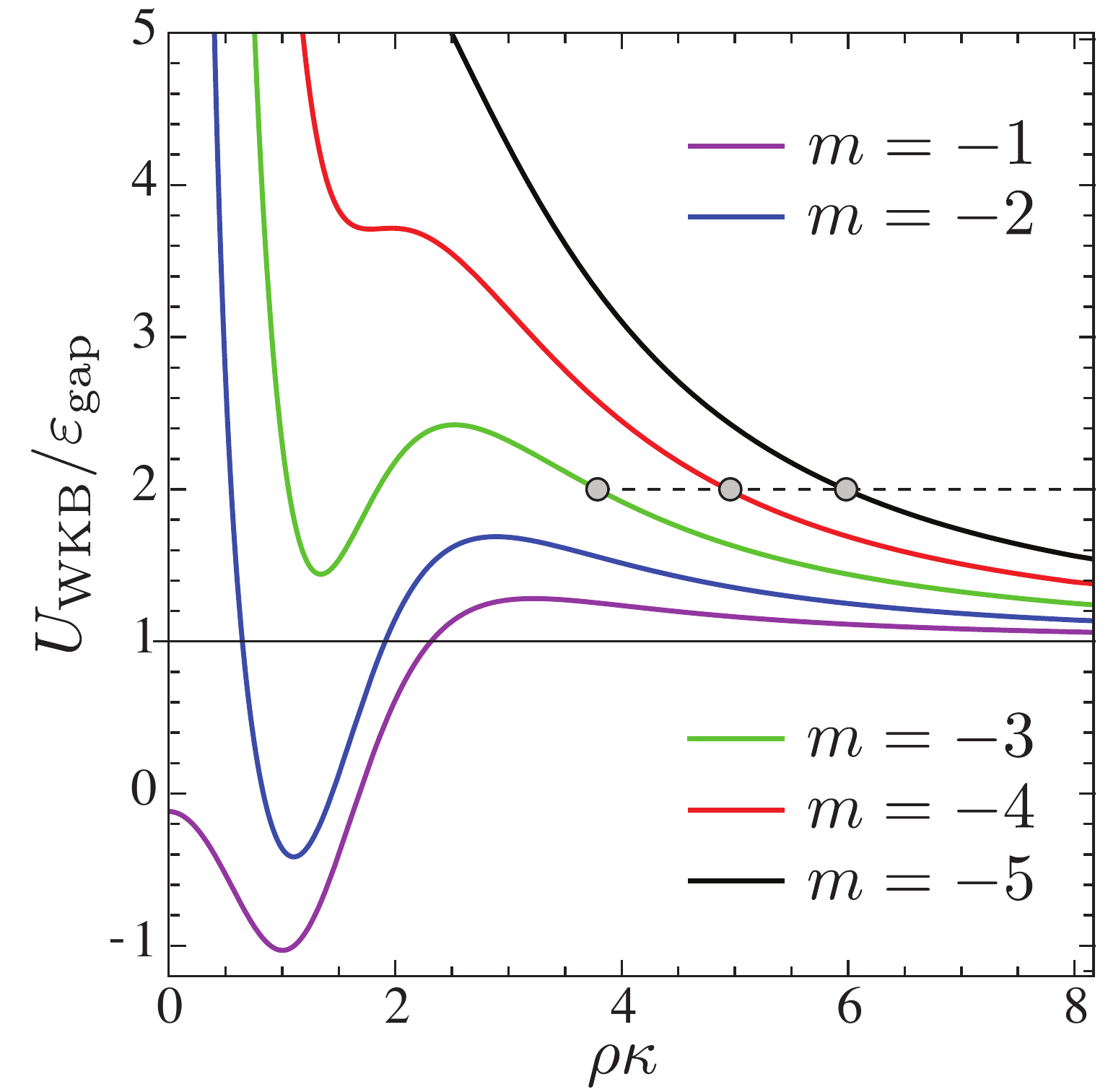}\quad
\includegraphics[width=0.48\columnwidth]{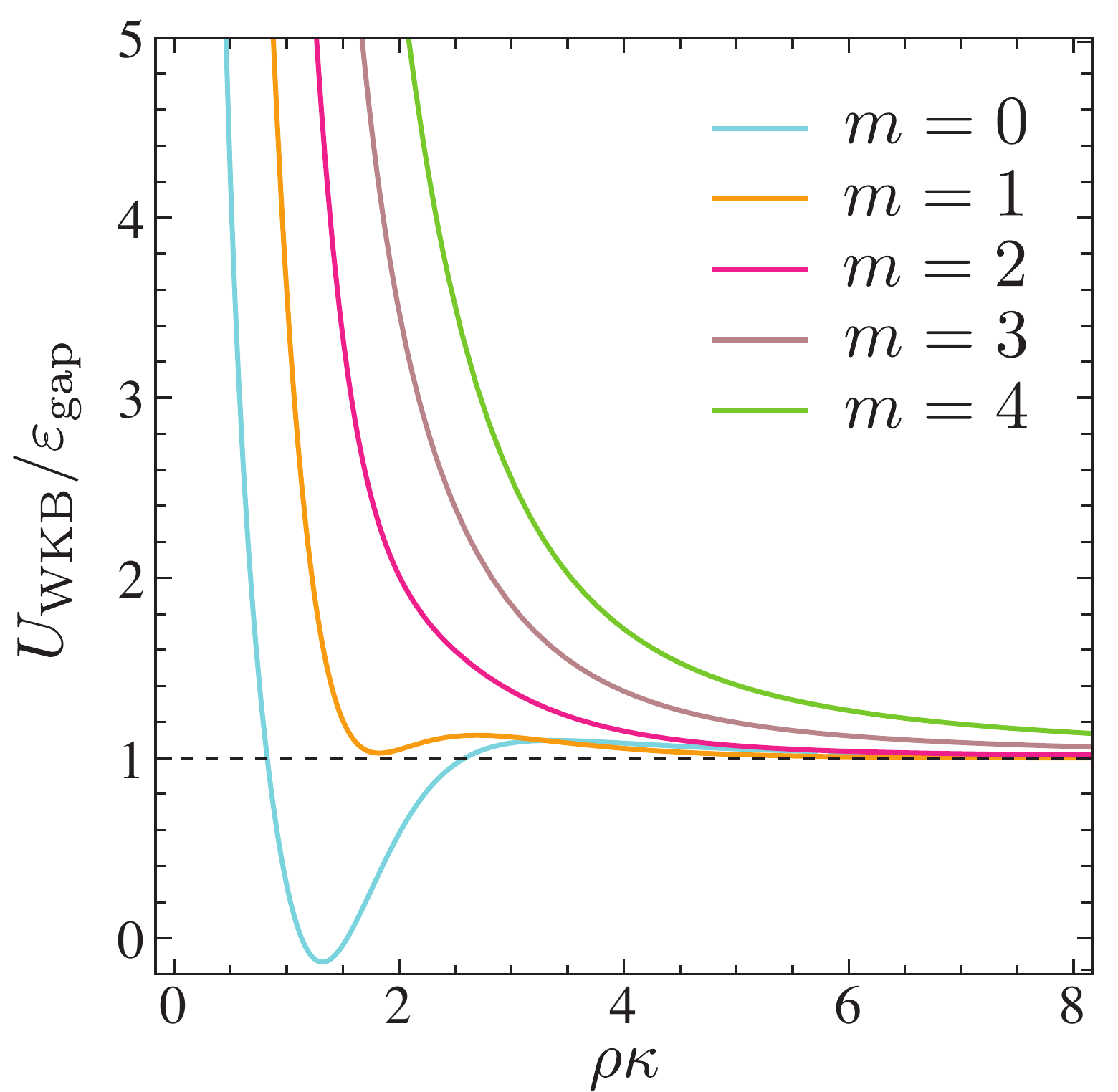}
\caption{Effective WKB potential $U_{\rm WKB}$, Eq.~\eqref{WKBpotential}, of the magnon scattering states for various values of angular momentum $m$ for $\kappa=Q$. In the classically allowed regimes, the energy of the scattering states obeys $\varepsilon > U_{\rm WKB}$. In the left panel, classical turning points, $\rho_0$, for the energy $\varepsilon = 2 \varepsilon_{\rm gap}$ are indicated by dots. For certain values of $m$ the potential develops a local maximum giving rise to resonances that are reflected in a pronounced energy dependence of the phase shift.} 
\label{fig:WKBpotential}
\end{figure}

For the discussion of the WKB approximation, we follow Langer \cite{Langer1937}, see also Ref.~\onlinecite{Berry1972}, and first substitute for the radius $\rho = e^x/\kappa$ and $\vec u(x) = \vec \eta_m(e^x/\kappa)$ so that the eigenvalue equation \eqref{EigenvalueEq} simplifies for any $x \in (-\infty,\infty)$ to 
\begin{align} \label{WKBForm}
\Big[-\partial_x^2 - \Lambda(x) \Big]  \vec u(x) = 0,
\end{align}
where the matrix $\Lambda$ is defined by 
\begin{align}
&\Lambda(x) = 
\\\nn&=
- (m\mathds{1} -\tau^z)^2 -
e^{2x} (\mathds{1}  - \tau^z \frac{\varepsilon}{\varepsilon_{\rm gap}} + \frac{1}{\varepsilon_{\rm gap}} \mathcal{V}_m(e^{x}/\kappa) ).
\end{align}
Eq.~\eqref{WKBForm} has the form of a two-component Schr\"odinger equation. The semiclassical approximation for such a problem is discussed for example in Ref.~\onlinecite{Frisk1993}. With the help of the WKB Ansatz $\vec u(x) = \vec u_{0}(x) e^{i S(x)}$ and neglecting derivatives of $\vec u_{0}(x)$, the above differential equation is converted into an algebraic one
\begin{align} 
\Big[(S'(x))^2 \mathds{1} - \Lambda(x) \Big]  \vec u_{0}(x) = 0.
\end{align}
This equation can be locally diagonalised for each position $x$. 
The interesting eigenvector is the one that becomes proportional to $\vec u_{0}^T(x) \propto (1 ,0)$ for large distances. The corresponding eigenvalue denoted as $\lambda(x)$ then determines the function $S'(x)$ with the help of which the phase of the wave function can be evaluated in the lowest order WKB approximation
\begin{align}
S(x) = \int^x dx' \sqrt{\lambda(x')} = \int^\rho d\rho' \frac{\sqrt{\lambda(\log(\kappa \rho'))}}{\rho'} .
\end{align}
The evolution of the eigenvectors with $x$ gives rise to Berry phases that however contribute only in next-to-leading order. The effective WKB potential
\begin{align} \label{WKBpotential}
U_{\rm WKB}(\rho) = \varepsilon - \varepsilon_{\rm gap} \frac{\lambda(\log(\kappa \rho))}{\rho^2\kappa^2},  
\end{align}
that is independent of the energy $\varepsilon$, is shown in Fig.~\ref{fig:WKBpotential}  for $\kappa = Q$. This potential possesses a single classical turning point for angular momenta $m \leq -5$ and $m \geq 2$. For other values of $m$, the potential develops a local maximum and for certain energies three classical turning points then appear. Neglecting corrections due to those additional classical turning points, the scattering phase shift in the WKB approximation is given by\cite{Berry1972}
\begin{align} \label{WKBPhaseShift}
\delta_m^{\rm WKB} &= 
\lim_{\tilde \rho \to \infty} \int_{\rho_0}^{\tilde \rho} 
\left( \sqrt{k^2 + \kappa^2 - \kappa^2 \frac{U_{\rm WKB}(\rho)}{\varepsilon_{\rm gap}}} - k \right) 
d\rho 
\nonumber\\ &
+ \frac{\pi}{2}|m-1|-k \rho_0,
\end{align}
where $k = \kappa \sqrt{\varepsilon/\varepsilon_{\rm gap} - 1}$. The distance $\rho_0$ corresponds here to the first classical turning point when approaching the potential from large distances.

The results for $\delta_m^{\rm WKB}$ are shown in Fig.~\ref{fig:phaseshift} as dashed lines. 
It provides a good approximation to the numerically exact values (solid lines) for high energies and 
higher angular momenta $|m|$. The sharp resonances for $m=-3$ and $m=-4$ are attributed to quasi bound states of the effective potential \eqref{WKBpotential}.

\subsubsection{Scattering cross section}

The scattering amplitude $f$ is defined in terms of the long-distance asymptotic behaviour of the magnon wave function but in the laboratory orthogonal frame as defined in Eq.~\eqref{LabFrame},  
\begin{align} \label{ScatteringPsi}
\vec \psi^{\rm scatter}_{\rm LAB}(\vect r - \vect R) = \left(\begin{array}{c} 1\\0 \end{array} \right) 
\Big(
e^{i \vect k (\vect r - \vect R)} + f(\chi) \frac{e^{i k \rho}}{\sqrt{\rho}} 
\Big)
\end{align}
where $(\vect r - \vect R)^T = \rho (\cos \chi, \sin \chi)$. Comparison with Eq.~\eqref{AsLarge} yields for the scattering amplitude $f$
\begin{align} \label{ScaAmplitude}
f(\chi) = \frac{e^{-i \frac{\pi}{4}}}{\sqrt{2 \pi k}} \sum^\infty_{m=-\infty} e^{i m \chi} \left(e^{i 2 \delta_{m+1}} - 1 \right),
\end{align}
with the momentum $k = \kappa \sqrt{\varepsilon/\varepsilon_{\rm gap} - 1}$. Note that the phase shift $\delta_{m+1}$ as defined in the local orthogonal frame \eqref{AsLarge} enters the sum with the angular momentum $m+1$. The differential scattering cross section is then given by
\begin{align} \label{diffscatcross}
\frac{d \sigma(\varepsilon)}{d\chi} = |f(\chi)|^2. 
\end{align}
The total scattering cross section finally reads
\begin{align}
\sigma(\varepsilon) = \frac{4}{k}\sum^\infty_{m=-\infty} \sin^2 \delta_{m+1}.
\end{align}

\begin{figure}
\centering
(a)\hfill {$\mbox{}$} \\
\includegraphics[width=0.8\columnwidth]{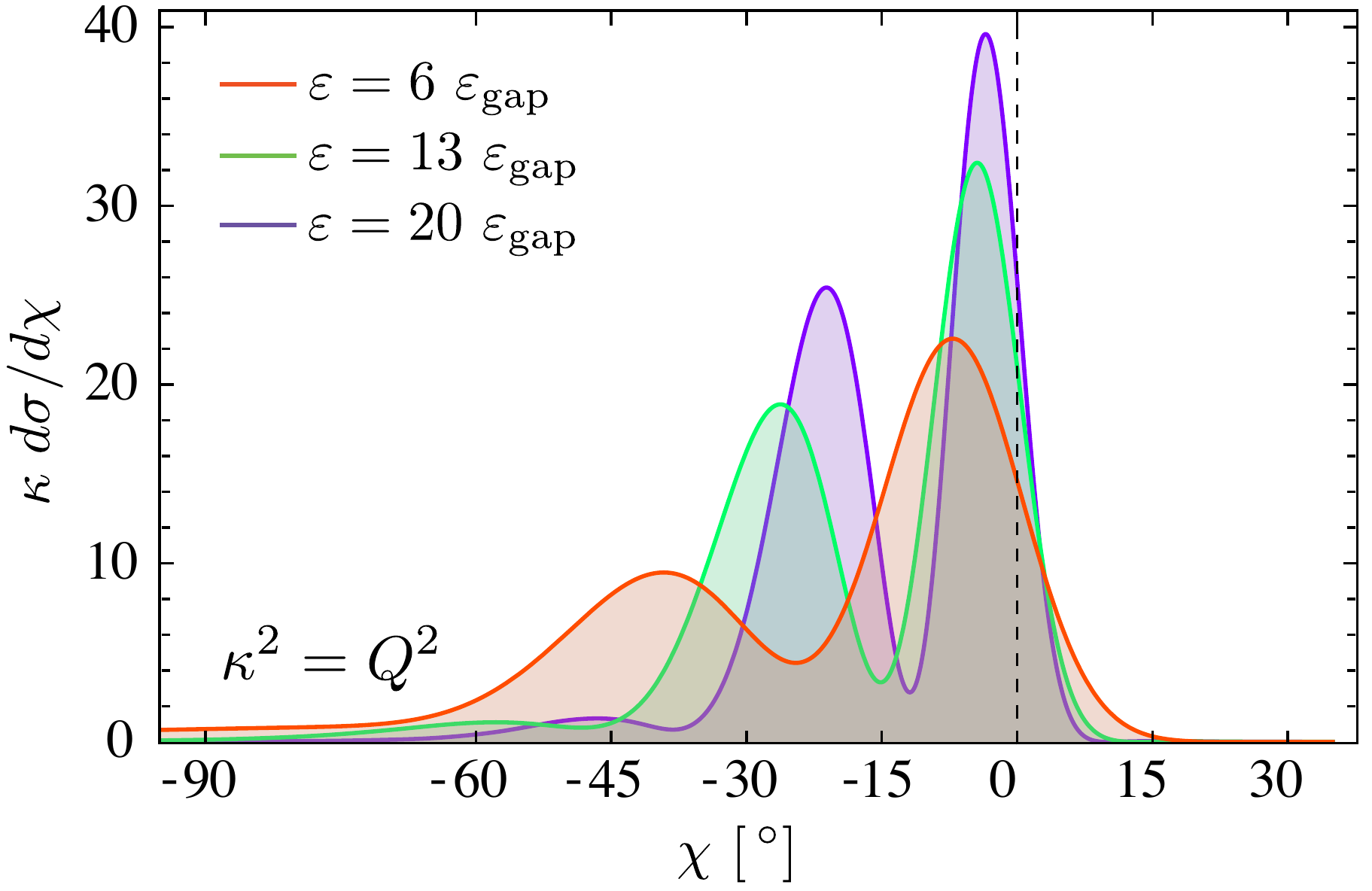}\\[1em]
(b)\hfill {$\mbox{}$} \\
\includegraphics[width=0.8\columnwidth]{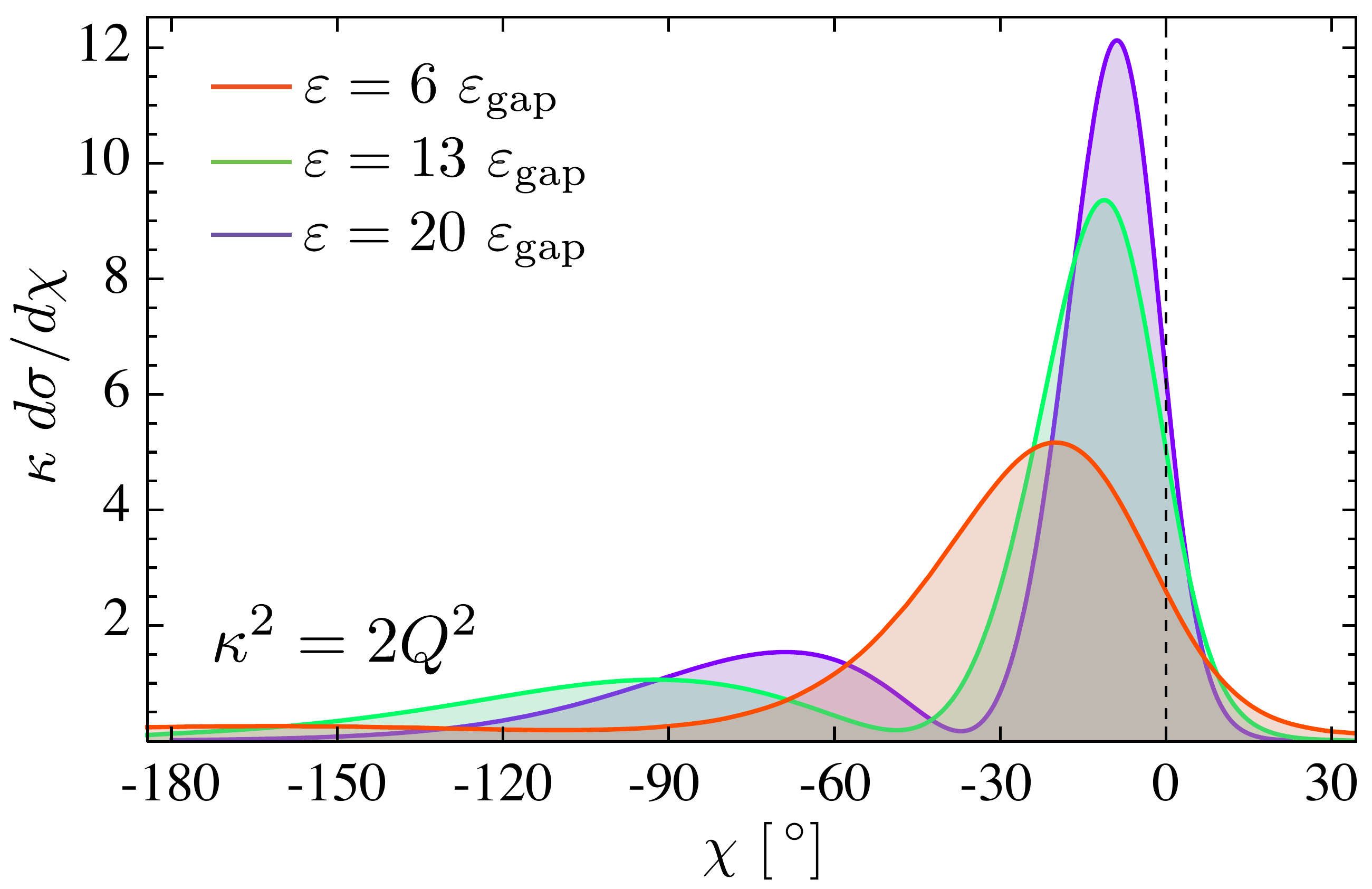}
\caption{Differential cross section \eqref{diffscatcross} evaluated within the WKB approximation for various energies $\varepsilon$ at magnetic fields (a) $\kappa^2 = Q^2$ and (b) $\kappa^2 = 2 Q^2$. Skew scattering results in a pronounced asymmetry with respect to forward scattering $\chi=0$ with  characteristic multiple peaks at high energies.
}
\label{fig:diffcrosssection}
\end{figure}

\begin{figure}
\centering
(a)\hfill {$\mbox{}$} \\
\includegraphics[width=0.85\columnwidth]{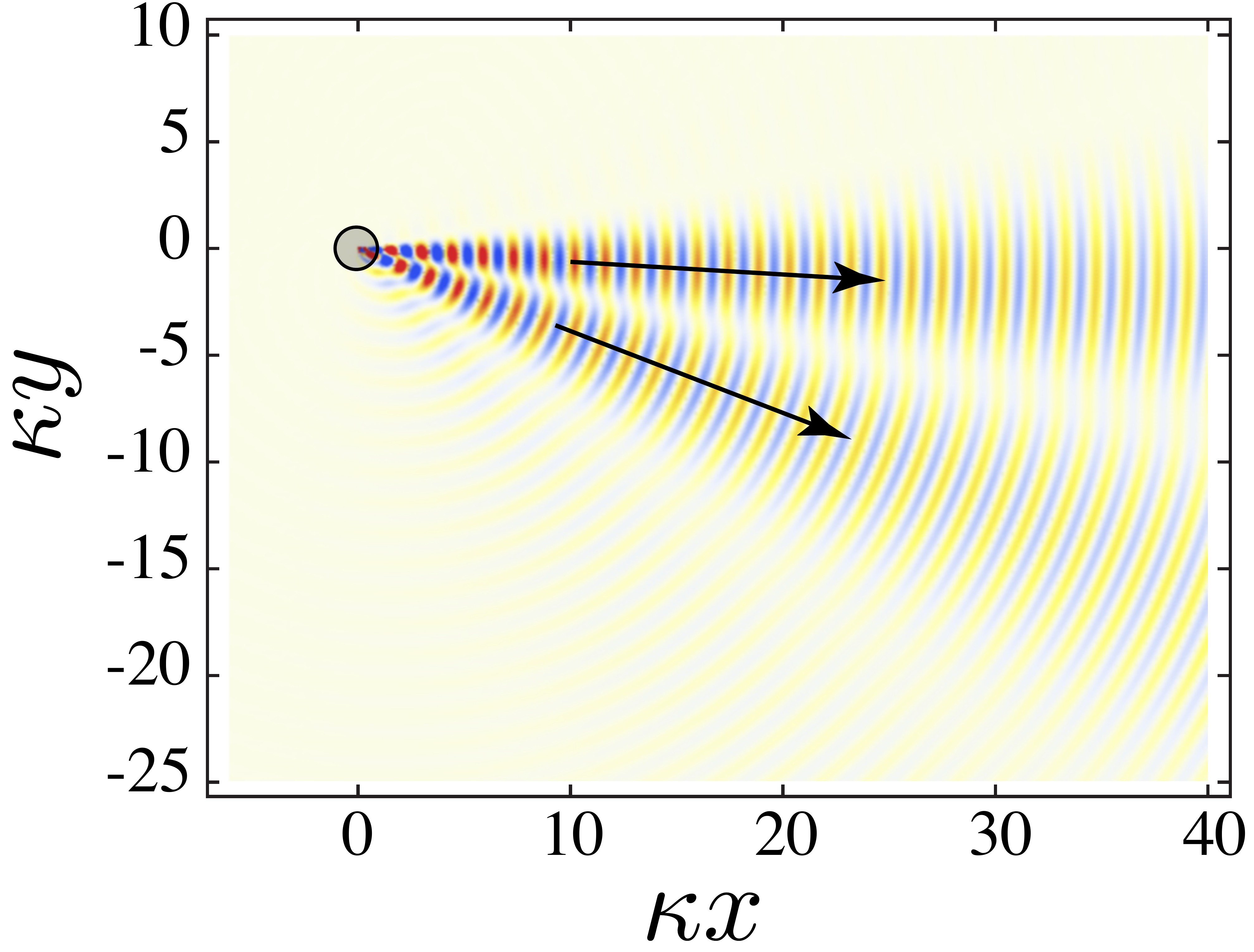}\\[1em]
(b)\hfill {$\mbox{}$} \\
\includegraphics[width=0.85\columnwidth]{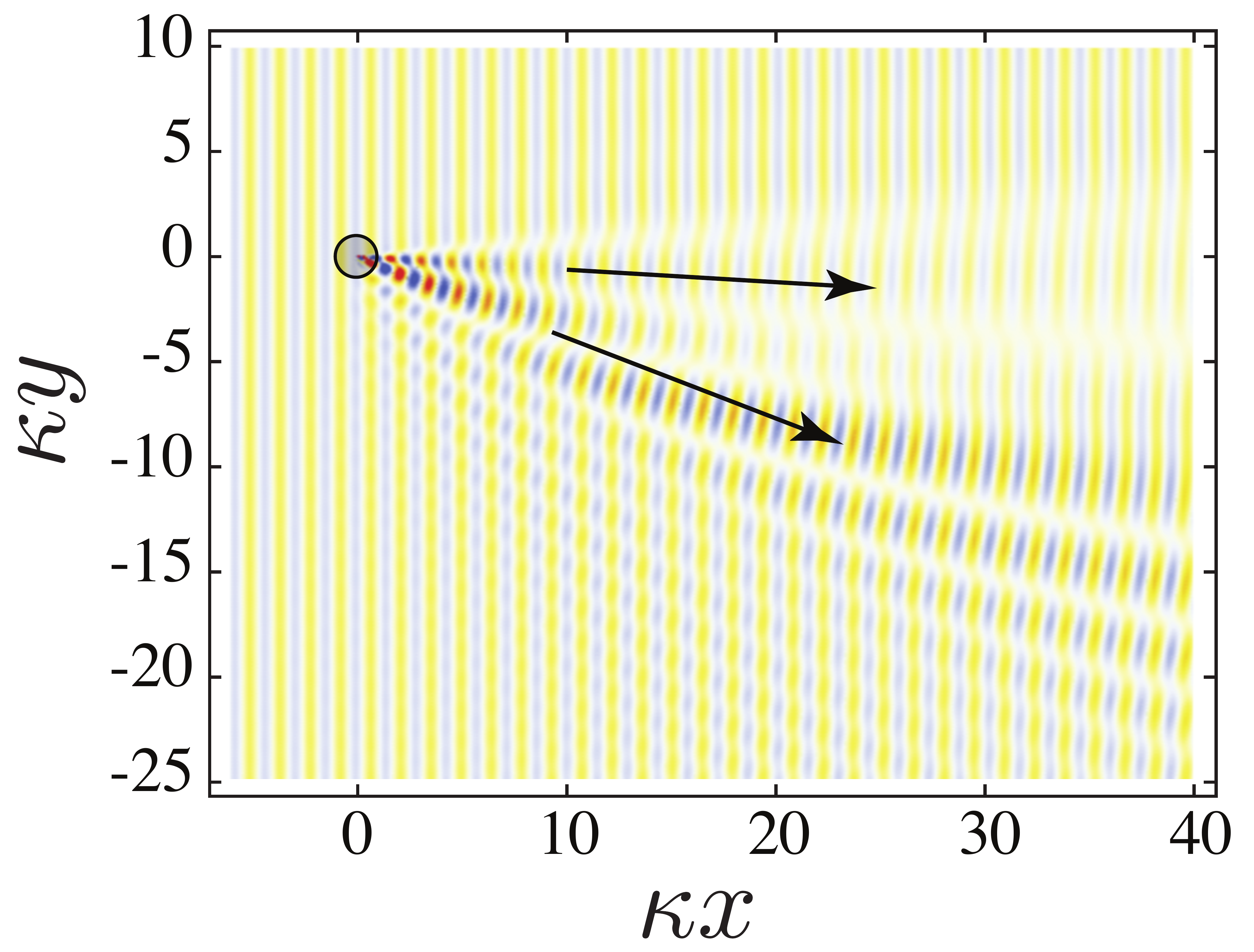}
\caption{Scattering wave function in the WKB approximation for the energy $\varepsilon = 20 \varepsilon_{\rm gap}$ at $\kappa = Q$ for an incoming wave with $\hat k = \hat x$. The arrows indicate the position of peaks in the differential scattering cross section, see Fig.~\ref{fig:diffcrosssection}. The skyrmion at the origin is represented by the circle with radius $1/\kappa$. Panel (a) shows only the scattered wave Re$\{(1,0)\vec \psi^{\rm scatter}_{\rm LAB}(\vect r) - e^{i \vect k \vect r}\}$ and panel (b) displays Re$\{(1,0)\vec \psi^{\rm scatter}_{\rm LAB}(\vect r)\}$ of Eq.~\eqref{ScatteringPsi} that exhibits the interference between the incoming and the scattered wave.}
\label{fig:WKBwavefunction}
\end{figure}

An evaluation of the differential scattering cross section within the WKB approximation is shown in Fig.~\ref{fig:diffcrosssection}. 
More precisely, we calculated the scattering amplitude \eqref{ScaAmplitude} entering \eqref{diffscatcross} using the WKB approximation for the phase shifts \eqref{WKBPhaseShift}, and the sum over angular momentum was cut off for $|m| > 30$. The cross section exhibits a pronounced asymmetry with respect to forward scattering, $\chi = 0$, with characteristic multiple peaks whose positions shift with energy. Note that a peak in $d\sigma/d\chi$ at a negative angle $\chi$ corresponds to a pronounced scattering in clockwise direction, i.e., to the right-hand side from the perspective of the incoming wave. This so-called {\it skew scattering} is also illustrated in Fig.~\ref{fig:WKBwavefunction} where we show the scattering wave function in the same WKB approximation.

We consider here only the elastic scattering of magnons. There are also inelastic scattering processes, for example, when the magnons excite the breathing bound state with $m=0$ in Fig.~\ref{fig:spectrum}. 
The inelastic scattering cross section of the skyrmion is  beyond the scope of the present work. 

\subsection{Discussion of the magnon-skyrmion scattering}
\label{sec:DiscScattering}

The skew scattering of magnons was recently observed in micromagnetic simulations by Iwasaki {\it et al.}\cite{Iwasaki2014}. These authors also present a theory for the differential cross section and explain that the skew scattering arises from an effective Lorentz force that emerges from the topological skyrmion texture. 
The theory of Ref.~\onlinecite{Iwasaki2014} for $d\sigma/d\chi$ differs from ours, however, in two aspects.  Firstly, the effective gauge potential for the magnons was assumed to decay algebraically in Ref.~\onlinecite{Iwasaki2014} giving rise to a finite total effective flux, whereas ours decreases exponentially with distance resulting in a vanishing total effective flux. Secondly, the resulting effective flux density that we obtain possesses a singularity at the skyrmion center. In the following, we explain in detail that this peculiar flux density profile not only leads to skew scattering but is also at the origin of {\it rainbow scattering} resulting in multiple peaks in the differential cross section of Fig.~\ref{fig:diffcrosssection}.

\subsubsection{Scattering from an effective magnetic flux}

\begin{figure}
\centering
\includegraphics[width=0.8\columnwidth]{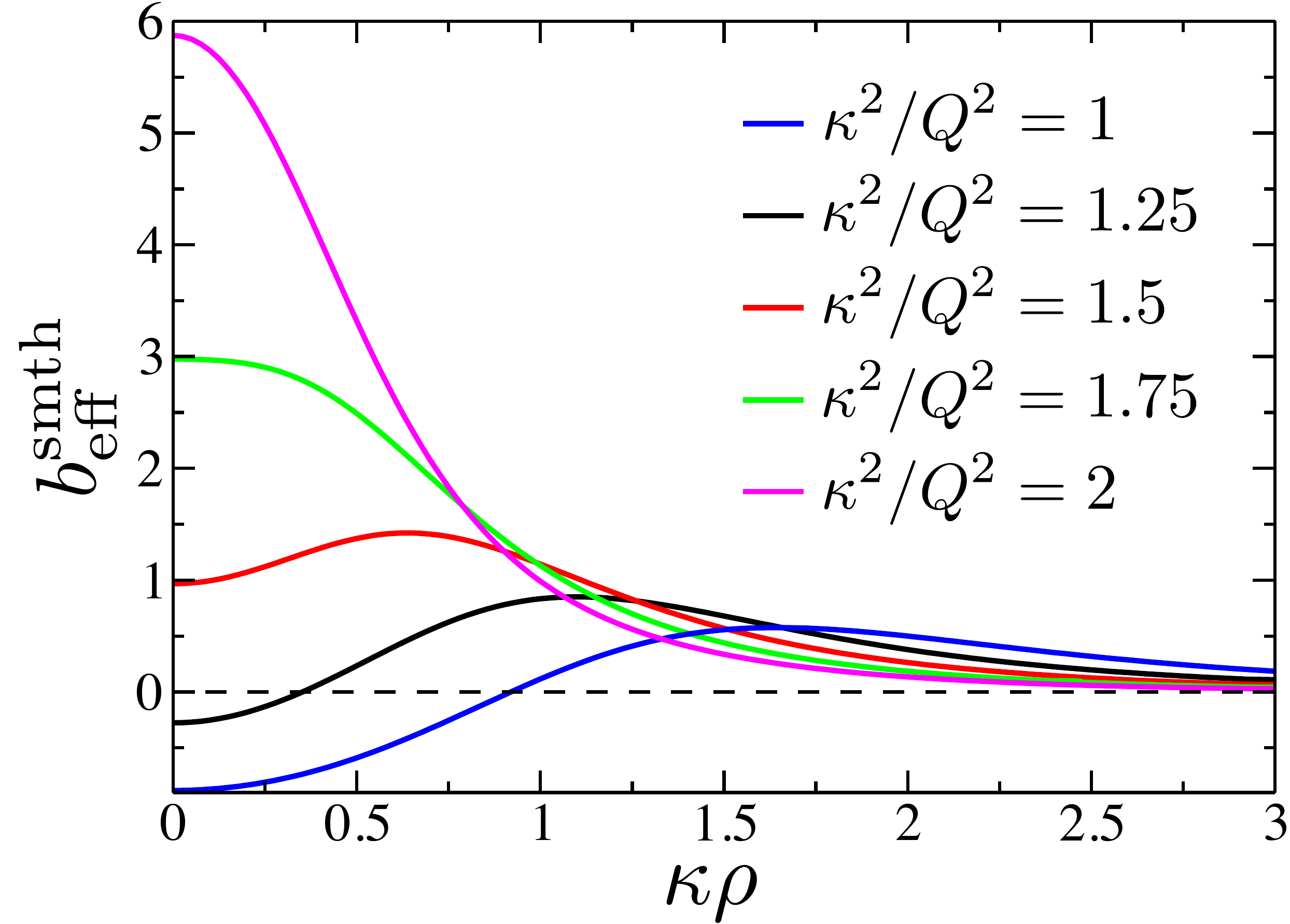}
\caption{Smooth part of the effective magnetic flux density \eqref{SmoothFlux} for various values of $\kappa$.
}
\label{fig:SmoothB}
\end{figure}

For high energies, $\varepsilon \gg \varepsilon_{\rm gap}$, the scattering potential is essentially governed by $v_z(\rho)$ of Eqs.~\eqref{ScatteringPots}, see Fig.~\ref{fig:potentials}, which in particular determines the position of the classical turning point. 
As $v_z(\rho)$ is proportional to the angular momentum $m$, this corresponds to scattering from an effective magnetic flux. Indeed, the effective gauge potential for the magnons within the laboratory orthogonal frame is obtained from Eq.~\eqref{GaugeField} with the help of the singular gauge transformation Eq.~\eqref{LabFrame}, $\psi \to \psi_{\rm LAB} = -i e^{- i\varphi} \psi$, and Eq.~\eqref{SkyrmionParameterization},
\begin{align} \label{LABGaugePotential}
\vec a_{\rm LAB}  &= \vec a - \nabla \chi = a_{\rm LAB}^\chi \hat \chi = 
\left(\frac{\cos \theta - 1}{\rho} - Q \sin \theta \right) \hat \chi,
\end{align}
and $v_z(\rho) = -2 m a_{\rm LAB}^\chi(\rho)/\rho$. This gauge potential $\vec a_{\rm LAB}$ is exponentially confined to the skyrmion area, which according to Stoke's theorem implies a vanishing total flux. The corresponding flux density, however, possesses an interesting structure. The first term on the right hand side of Eq.~\eqref{LABGaugePotential} diverges for small radii as $-2/\rho$ for $\theta \to \pi$ giving rise to a singular flux density,
\begin{align} \label{MagnonFluxDensity}
b_{\rm eff}(\vect r - \vect R) \equiv \varepsilon_{z\alpha \beta} \partial_\alpha \vec a_{{\rm LAB},\beta} = -4\pi \delta(\vect r - \vect R) + b^{\rm smth}_{\rm eff}(\rho).
\end{align}
The smooth part reads explicitly
\begin{align} \label{SmoothFlux}
&b^{\rm smth}_{\rm eff}(\rho) = 
\\\nn& \partial_\rho \left(\frac{\cos \theta -1}{\rho} - Q \sin \theta \right) + \frac{1}{\rho}\left(\frac{\cos \theta -1}{\rho} - Q \sin \theta \right).
\end{align}
It integrates to 
\begin{align}
2\pi \int_0^\infty d\rho\, \rho\, b^{\rm smth}_{\rm eff}(\rho) = 2\pi(\cos \theta - Q \rho \sin \theta) \Big|^{\infty}_{0} = 4\pi
\end{align}
and thus exactly cancels the singular part. The dependence of $b^{\rm smth}_{\rm eff}(\rho)$ on the radius $\rho$ is shown in Fig.~\ref{fig:SmoothB} for various values of $\kappa$. For $\kappa^2/Q^2 \lesssim 1.7$, it possesses a local maximum, and it even changes sign as a function of $\rho$ for $\kappa^2/Q^2 \lesssim 1.3$.

Why is the flux density singular? Consider the orthogonal frame after the singular gauge transformation \eqref{LabFrame}, see Eq.~\eqref{GaugeTrafo}, e.g., $\hat e^+_{\rm LAB} = i e^{i \varphi} \hat e^+$, as a function of distance from the skyrmion center. For large distances $\theta \to 0$, and we recover $\hat e^+_{\rm LAB} = \frac{1}{\sqrt{2}}(\hat x + i \hat y)$ for $\theta = 0$. On the other hand, very close to the skyrmion center $\theta \approx \pi$, and $\hat e^+_{\rm LAB} \approx \frac{1}{\sqrt{2}}(\hat x - i \hat y) e^{i 2 \chi}$ becomes dependent on the polar angle $\chi$. It corresponds to an effectively rotating frame that rotates twice upon encircling the core once, resulting in the singular flux of $- 4\pi$ in Eq.~\eqref{MagnonFluxDensity}.

Keeping only the effective magnetic scattering potential, the WKB potential of Eq.~\eqref{WKBpotential} simplifies to 
\begin{align}
\frac{U_{\rm WKB}(\rho)}{\varepsilon_{\rm gap}} \approx  1 + 
\frac{(m-1)^2-2m  \rho a^\chi_{\rm LAB}(\rho)}{\kappa^2\rho^2}. 
\end{align}
At high energies, the classical turning point of this potential is asymptotically determined by the limiting value $\rho a^\chi_{\rm LAB}(\rho) \to -2 $ for $\rho \to 0$ corresponding to the scattering off a singular magnetic string with flux $-4\pi$. In this limit, the WKB scattering phase shift \eqref{WKBPhaseShift} assumes the Aharonov-Bohm form,\cite{Aharonov1959} $\delta^{\rm WKB}_m \to \delta^{\rm AB}_m$ with
\begin{align} \label{AB}
\delta_{m}^{\rm AB} &= - \frac{\pi}{2} |m+1|
+ \frac{\pi}{2}|m-1|
= \left\{
\begin{array}{lll}
-\pi & {\rm if} & m > 0 \\
0& {\rm if} & m = 0 \\
\pi & {\rm if} & m < 0
\end{array}
\right..
\end{align}
This explains the asymptotic values obtained for the phase shifts in Fig.~\ref{fig:phaseshift} in the limit of high energies.

In the limit of low energies, $\varepsilon \to \varepsilon_{\rm gap}$, on the other hand, one expects the phase shifts to obey Levinson's theorem generalized to the case of Aharonov-Bohm scattering,\cite{Sheka2006}
\begin{align}
\delta_m(\varepsilon_{\rm gap}) - \delta_m(\infty) = \pi N^{\rm bound}_m - \delta_m^{\rm AB}, 
\end{align}
where $N^{\rm bound}_m$ is the number of bound states with angular momentum number $m$. For our definition of the phase shift $\delta_m(\infty) = \delta_m^{\rm AB}$, and one obtains $\delta_m(\varepsilon_{\rm gap})/\pi = N^{\rm bound}_m$. For the specific value $\kappa^2 = Q^2$ there are two magnon bound states present, see Fig.~\ref{fig:spectrum}: the breathing mode with $m=0$ and the zero mode with $m = - 1$. Consequently, one expects all phase shifts to vanish at the threshold $\varepsilon_{\rm gap}$ except $\delta_0(\varepsilon_{\rm gap}) = \delta_{- 1}(\varepsilon_{\rm gap}) = \pi$ for $\kappa^2 = Q^2$. 
This is in agreement with the numerically obtained values for the phase shifts presented in Fig.~\ref{fig:phaseshift}. 
%
Furthermore, the sharp drop of $\delta_{-2}(\varepsilon)$ close to the magnon gap can be attributed to the bound state with $m=-2$ that materializes for slightly smaller values of $\kappa$.

\subsubsection{Classical deflection function \& rainbow scattering}

\begin{figure}
\centering
\includegraphics[width=0.95\columnwidth]{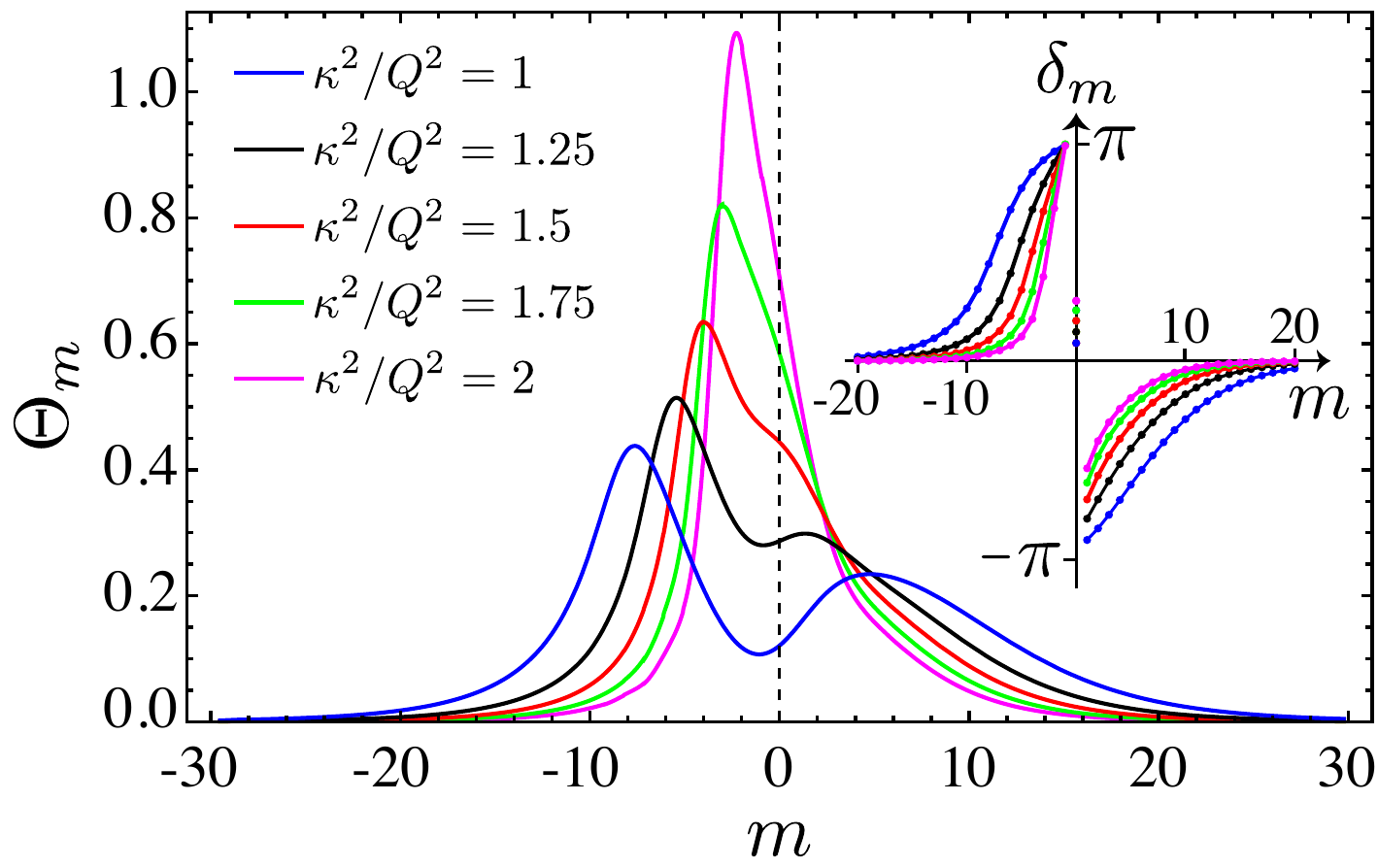}
\caption{Classical deflection function $\Theta_m = 2 \delta'_m$ of Eq.~\eqref{DeflectionFunction}
for the energy $\varepsilon = 20 \varepsilon_{\rm gap}$ for various values of $\kappa$. As a function of increasing $\kappa$, the three stationary points merge into a single maximum. The inset shows the corresponding phase shifts evaluated within the WKB approximation. 
}
\label{fig:ClassDefl}
\end{figure}

At fixed energies but in the limit of large angular momentum $|m| \gg 1$, i.e., large impact parameter, the phase shifts $\delta_m$ are eventually expected to vanish. Combined with the Aharanov-Bohm constraint, Eq.~\eqref{AB}, this has the consequence that $\delta_m$ as a function of increasing but negative $m$ has to increase towards $\pi$, and for positive and increasing $m$ it again must increase towards zero, see inset of Fig.~\ref{fig:ClassDefl}. Treating the phase shift $\delta_m$ as a continuous function of $m$ (except close to $m=0$), it follows that this function should change curvature. Equivalently, its derivative, the classical deflection function\cite{Ford1959,Herb1999,Fink2009} 
\begin{align} \label{DeflectionFunction}
\Theta_m = 2 \frac{\partial \delta_m}{\partial m},
\end{align}
possesses at least one stationary point with $\Theta'_m = 0$. This is illustrated in Fig.~\ref{fig:ClassDefl}. Note that the positive values obtained for $\Theta_m$ translate to a skew scattering at a mathematically negative angle $\chi$, that labels the vertical axes of Fig.~\ref{fig:diffcrosssection}. 

Generally, at such a stationary point multiple classical trajectories contribute to the scattering cross section resulting in so-called {\it rainbow scattering}.\cite{Ford1959} For a single stationary point, the scattering amplitude in the classical limit is then described by the Airy function resulting in a differential cross section with a sharp fall-off on the dark side and an oscillatory behavior on the bright side of the rainbow angle $\Theta_{m_{\rm st}}$ for which $\Theta'_{m_{\rm st}} = 0$.
Interestingly, the deflection function in Fig.~\ref{fig:ClassDefl} exhibits for $\kappa^2 = Q^2$ three stationary points whose contributions will interfere in $d\sigma/d\chi$ as shown in Fig.~\ref{fig:diffcrosssection}(a) with a weight determined by their curvature $\Theta''_{m}$. For increasing $\kappa$, the three stationary points merge into a single maximum, which in the classical limit governs  $d\sigma/d\chi$ at $\kappa^2 = 2 Q^2 $ as shown in Fig.~\ref{fig:diffcrosssection}(b).
The change in the number of stationary points of $\Theta_m$ is related to the smooth part of the effective flux density, see Fig.~\ref{fig:SmoothB}, which substantially alters its profile on a similar scale of $\kappa$.

\section{Magnon pressure on the skyrmion}
\label{sec:MagnonPressure}

Consider a plane wave of magnons impinging on the skyrmion as in Fig.~\ref{fig:WKBwavefunction}(b). What is the pressure on the skyrmion and will it be moving with a finite velocity $\dot{\vect R}$? 

A skyrmion motion in a corresponding numerical experiment was recently observed by Iwasaki {\it et al.}.\cite{Iwasaki2014} It was suggested by these authors that this motion can be explained in terms of total momentum conservation implying, in particular, that the skyrmion can be considered as a particle with well-defined momentum. 
However, the notion of a conserved momentum for the field theory \eqref{action} is subtle.\cite{Haldane1986,Volovik1987,Papanicolaou1990,Yan2013} We have recognized in section \ref{subsubsec:ZerothOrderInPsi} that in zeroth order in the massive modes $\psi$ the skyrmion coordinate, $\vect R$, obeys the equation of motion of a massless particle in a magnetic field, Eq.~\eqref{ClassicalEoM}. Its canonical momentum \eqref{SkyrmionMomentum} is spin-gauge dependent, and in general neither well-defined nor conserved. 
Nevertheless, it was argued in Ref.~\onlinecite{Papanicolaou1990} that the coordinates $\vect R_\alpha$ with $\alpha=1,2$ are conjugate to each other and, therefore, can be interpreted as a momentum of the skyrmion texture, as further discussed in Appendix \ref{app:Noether}.

We explain below that magnons indeed transfer momentum to the skyrmion. Using the conservation law associated with translation invariance, we show that a magnon current gives rise to a magnon pressure in the form of a momentum-transfer force on the skyrmion. This force enters the Thiele equation of motion for the skyrmion, that can be interpreted as a constant flow of momentum from the magnons to the skyrmion leading to a constant skyrmion velocity. Evaluating the corresponding force explicitly, we arrive at an expression for the  skyrmion velocity $\dot{\vect R}$ and the skyrmion Hall angle $\Phi$.

\subsection{Effective skyrmion equation of motion}
 
\subsubsection{Effective Thiele equation}

As shown in appendix \ref{app:Noether}, the conservation law deriving from space-time translation invariance reads,
see Eq.~\eqref{EMConservation1},
\begin{align} \label{EMConservation1again}
d_\mu T^{\rm stat}_{\mu\nu} = \frac{4\pi\hbar}{a^2} \epsilon_{\alpha 0 \nu} j_\alpha^{\rm top},
\end{align}
where $T^{\rm stat}_{\mu\nu}$ is the energy-momentum tensor obtained from the static part of the Lagrangian only, see Eq.~\eqref{EMTensor}, and $j_\alpha^{\rm top}$ is the spatial part of the topological current as defined in Eq.~\eqref{TopCurrent}. 

Expanding the topological current up to second order in the magnon fields we obtain
\begin{align}
&j^{\rm top}_\alpha =
\\\nn&j^{\rm top (0)}_0 d_0 \vect R_\alpha 
- \frac{1}{8\pi} \epsilon_{\alpha 0 \beta} 
\Big[d_0 (\vec \psi^\dagger \Gamma^\beta \vec \psi) + d_\beta (\vec \psi^\dagger \tau^z i d_0 \vec \psi) \Big] 
\end{align}
with the topological charge density of the static skyrmion
\begin{align}
j^{\rm top (0)}_0 = \frac{1}{4\pi} \hat n_s (\partial_1 \hat n_s \times \partial_2 \hat n_s) = \frac{1}{4\pi} \frac{\theta' \sin \theta}{\rho}.
\end{align}
The vertex $\Gamma^\beta$ is just the interaction vertex of Eq.~\eqref{InteractionVertex}. Integrating the spatial component of Eq.~\eqref{EMConservation1again} over space, we arrive at an equation of motion for the skyrmion given at this order by 
\begin{align} \label{MagnonPressureEoM}
\vect G \times d_0 \vect R = \vect F,
\end{align}
with $\vect G = -\frac{4\pi\hbar}{a^2} \hat z$. It is just the Thiele equation of Eq.~\eqref{ClassicalEoM} but in the presence of an additional force $\vect F$, that is given by 
\begin{align} \label{MagnonPressureForce}
&\vect F_\alpha = 
\\\nn
&- \int d^2 \vect r \Big[d_\beta T^{\rm stat}_{\beta \alpha} + \frac{\hbar}{2a^2}  \Big(d_0 (\vec \psi^\dagger \Gamma^\alpha \vec \psi) + d_\alpha( \vec \psi^\dagger \tau^z i d_0 \vec \psi) \Big)\Big].
\end{align}
We would like to determine the skyrmion velocity only in linear response, so that we can limit ourselves to evaluate the force $\vect F$ in zeroth order in $d_0 \vect R$. 
In a stationary 
scattering situation the second term on the right-hand side in Eq.~\eqref{MagnonPressureForce} vanishes and we neglect it in the following. The integrand then reduces to a total derivative, and the force is given by a surface integral. Choosing the surface to be a circle with radius $\rho_L \ggg 1/\kappa$ centered around the skyrmion we finally obtain
\begin{align} \label{MagnonForce1}
\vect F_\alpha &= 
- \rho_L \int_0^{2\pi} d\chi 
\Big[\hat \rho_\beta T^{\rm stat}_{\beta \alpha} + \frac{\hbar}{2a^2} \hat \rho_\alpha (\vec \psi^\dagger \tau^z i d_0 \vec \psi)\Big]_{\rho_L} 
\nn\\ &= - \rho_L \int\limits_0^{2\pi} d\chi\, \hat \rho_\beta T^{\rm mag}_{\beta \alpha} \Big|_{\rho_L}
\end{align}
One can verify that for large radii $\rho_L$ the integrand is just given by the spatial component of the energy-momentum tensor of  free magnons,
\begin{align} 
T^{\rm mag}_{\mu\nu} = \frac{\partial \mathcal{L}_{\rm mag}}{\partial (d_\mu \vec \psi_{\rm LAB})} d_\nu \vec \psi_{\rm LAB} + h.c. 
- \delta_{\mu \nu} \mathcal{L}_{\rm mag},
\end{align}
with the Lagrangian
\begin{align}  \label{MagLagrangian}
&\mathcal{L}_{\rm mag} =
\frac{1}{2 a^2} \Big\{ 
\frac{\hbar}{2} \Big(\vec \psi^\dagger_{\rm LAB} \tau^z   d_\tau \vec \psi_{\rm LAB} - (d_\tau \vec \psi^\dagger_{\rm LAB}) \tau^z   \vec \psi_{\rm LAB}\Big)
\nn\\&\quad+ \frac{\hbar^2}{2M_{\rm mag}}
(d_\alpha \vec \psi^\dagger_{\rm LAB})(d_\alpha \vec \psi_{\rm LAB})
+ \kappa^2 \vec \psi^\dagger_{\rm LAB} \vec \psi_{\rm LAB}
\Big\}
\end{align}
but with the magnon field defined within the laboratory orthogonal frame, see Eq.~\eqref{LabFrame}, and we have used 
the expression for the magnon mass $M_{\rm mag} = \hbar^2/(2\varepsilon_0 a^2)$.

So we arrive at the result that the force, $\vect F$, is just determined by the net current of momentum, $T^{\rm mag}_{\beta \alpha}$, carried by the magnons through the surface of the sample. This net momentum is transfered to the skyrmion. Indeed, using the results of appendix \ref{app:Mom&AngMom} we can associate with the skyrmion a momentum $\vect P^{\rm skyr}_\alpha = \frac{4\pi\hbar}{a^2} \epsilon_{0 \alpha \beta} \vect R_\beta$.\cite{Papanicolaou1990,Komineas1996,Komineas2008} The Thiele equation \eqref{MagnonPressureEoM} can then be rewritten in the following form
\begin{align} \label{MagnonPressureEoM2}
d_0 \vect P^{\rm skyr} = \vect F,
\end{align}
which explicitly describes the flow of momentum between the magnon subsystem and the skyrmion.

\subsubsection{Momentum-transfer force on the skyrmion}

The force \eqref{MagnonForce1} only depends on the magnon wave function far away from the skyrmion so that we can use its asymptotic scattering form. 
We consider a magnon plane wave impinging from the left-hand side with fixed wave vector $\vect k = k \hat x$ and on-shell energy $\varepsilon$
\begin{align} \label{PlaneMagnonWave}
\vec \psi_{\rm LAB}(\vect r - \vect R,t) = \sqrt{\delta}\, e^{-i \varepsilon t/\hbar} \vec \psi^{\rm scatter}_{\rm LAB}(\vect r - \vect R),
\end{align}
as a function of real time $t = -i  \tau$. The amplitude is $\sqrt{\delta}$ and $\vec \psi^{\rm scatter}_{\rm LAB}$ was specified already in Eq.~\eqref{ScatteringPsi}. For the on-shell wavefunction $\vec \psi_{\rm LAB}$ the Lagrangian $\mathcal{L}_{\rm mag}$ vanishes, and the force is  determined by
\begin{align}
&\vect F_\alpha = \\\nn 
&- \rho_L \int_0^{2\pi} d\chi\, 
\hat \rho_\beta 
\frac{1}{2a^2}\frac{\hbar^2}{2M_{\rm mag}}
\Big[
(\partial_\beta \vec \psi^\dagger_{\rm LAB})(\partial_\alpha \vec \psi_{\rm LAB}) + h.c.
 \Big]_{\rho_L}.
\end{align}
With the help of the optical theorem
\begin{align}
\sigma = 2 \sqrt{\frac{2\pi}{k}} {\rm Im}\{-i e^{i \pi/4} f(\chi=0)\}
\end{align}
we find for the force the explicit expression
\begin{align} \label{MagnonPressure}
\vect F = 
\frac{\delta}{a^2}\frac{\hbar^2k^2}{2M_{\rm mag}}
\left(\begin{array}{c} \sigma_\parallel(\varepsilon) \\ \sigma_\perp(\varepsilon) \end{array}  \right).
\end{align}
It depends on the energy-dependent longitudinal and transversal cross sections
\begin{align}
\left(\begin{array}{c} \sigma_\parallel(\varepsilon) \\ \sigma_\perp(\varepsilon) \end{array}  \right)
=
\int_0^{2\pi} d\chi \left(\begin{array}{c} 1- \cos\chi \\ -\sin \chi \end{array}  \right) \frac{d\sigma(\varepsilon)}{d\chi} .
\end{align}
The force along the magnon wave vector $\hat k^T = (1,0)$ is determined by $\sigma_\parallel$, which is an angular integral that weights the factor $1 - \cos \chi$ with the differential cross section. This structure is familiar from transport theory, and signals that the flow of magnon momentum remains unaltered for forward scattering resulting in a vanishing contribution to the force. The component of $\vect F$ perpendicular to the wave vector, on the other hand, is governed by $\sigma_\perp$ and is here only finite due to the asymmetric skew scattering. 

\subsection{Skyrmion velocity and Hall angle $\Phi$}

\begin{figure}
\centering
\includegraphics[width=0.75\columnwidth]{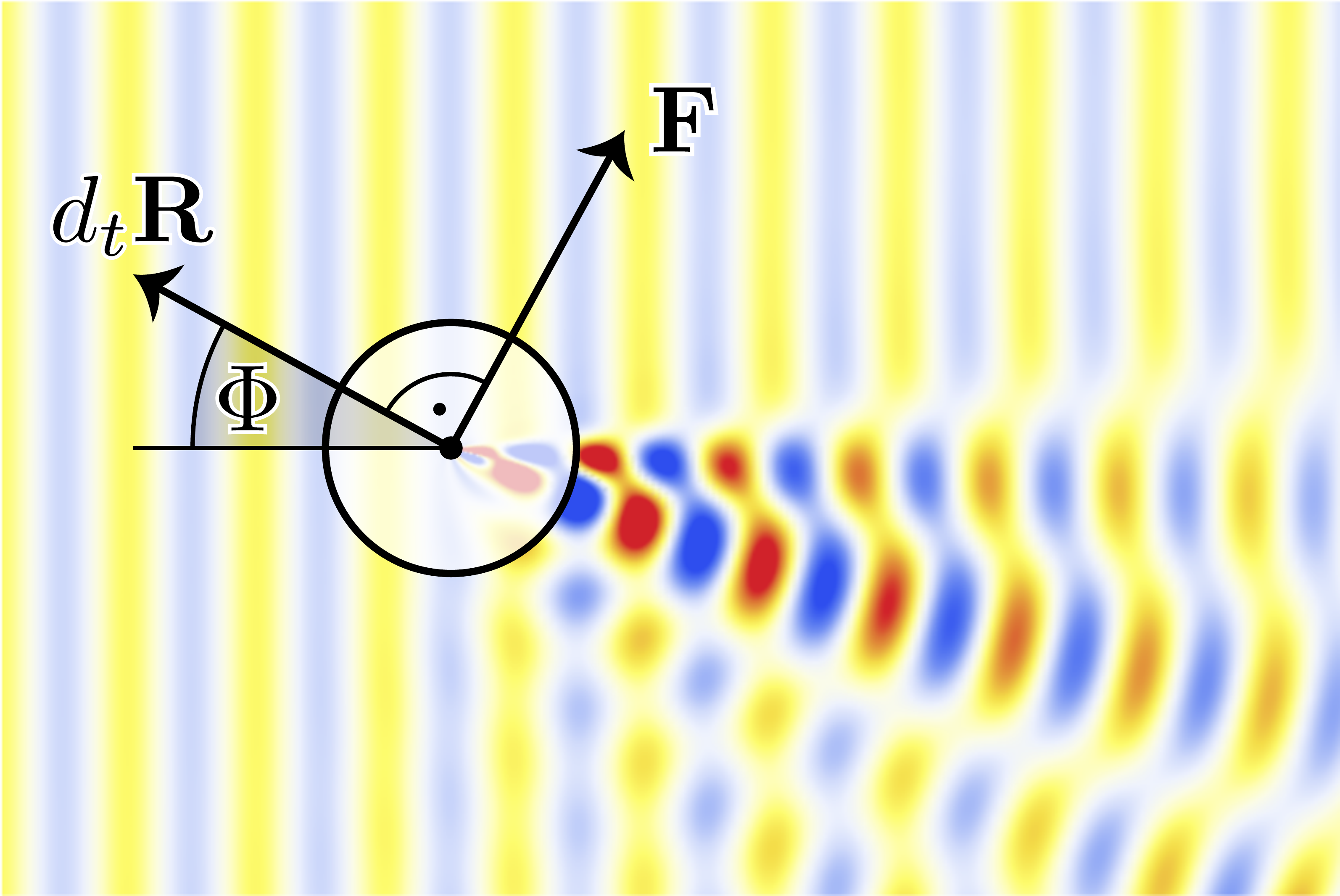}
\caption{Illustration of the magnon pressure and the resulting skyrmion velocity, $d_t \vect R$. The magnons are emitted from a source on the left-hand side, see also Fig.~\ref{fig:WKBwavefunction}, and scatter from the skyrmion represented by the circle with radius $1/\kappa$. The skyrmion experiences the force $\vect F$ given in Eq.~\eqref{MagnonPressure} which results in a finite skyrmion velocity $d_t \vect R$ of Eq.~\eqref{SkyrmionVelocity}, with a skyrmion Hall angle $\Phi$ given in Eq.~\eqref{SkyrmionHallAngle}. The arrows only illustrate the orientation but not the length of the corresponding vectors. 
}
\label{fig:magnonpressure}
\end{figure}

The equation of motion \eqref{MagnonPressureEoM} is easily solved and we obtain a constant skyrmion velocity $d_t \vect R = d_0 \vect R$ given by
\begin{align}
d_t \vect R_\alpha = -\frac{a^2}{4\pi\hbar} \epsilon_{0 \alpha \beta} \vect F_\beta.
\end{align}
The skyrmion velocity reads explicitly
\begin{align}  \label{SkyrmionVelocity}
d_t \vect R =  \frac{\delta}{8\pi} 
\frac{\hbar k^2}{M_{\rm mag}}
\left(\begin{array}{c} -\sigma_\perp(\varepsilon) \\ \sigma_\parallel(\varepsilon) \end{array}  \right).
\end{align}
The skyrmion velocity depends on its differential cross section, and its orientation is governed by $\sigma_\perp$ and $\sigma_\parallel$. 
In the linear response approximation, we can use the differential cross section evaluated in zeroth order in the velocity-interaction \eqref{MagnonVelocityInteraction} and can apply the results of the previous section for $d\sigma/d\chi$. 
At higher energies, the skyrmion velocity $d_t \vect R$ is approximately opposed to the direction into which the magnons are preferentially scattered as $\sigma_\parallel(\varepsilon) > 0$ and $\sigma_\perp(\varepsilon) > 0$, see the illustration in Fig.~\ref{fig:magnonpressure}. Such a direction for the skyrmion velocity was numerically observed in Ref.~\onlinecite{Iwasaki2014}.

\begin{figure}
\centering
\includegraphics[width=0.95\columnwidth]{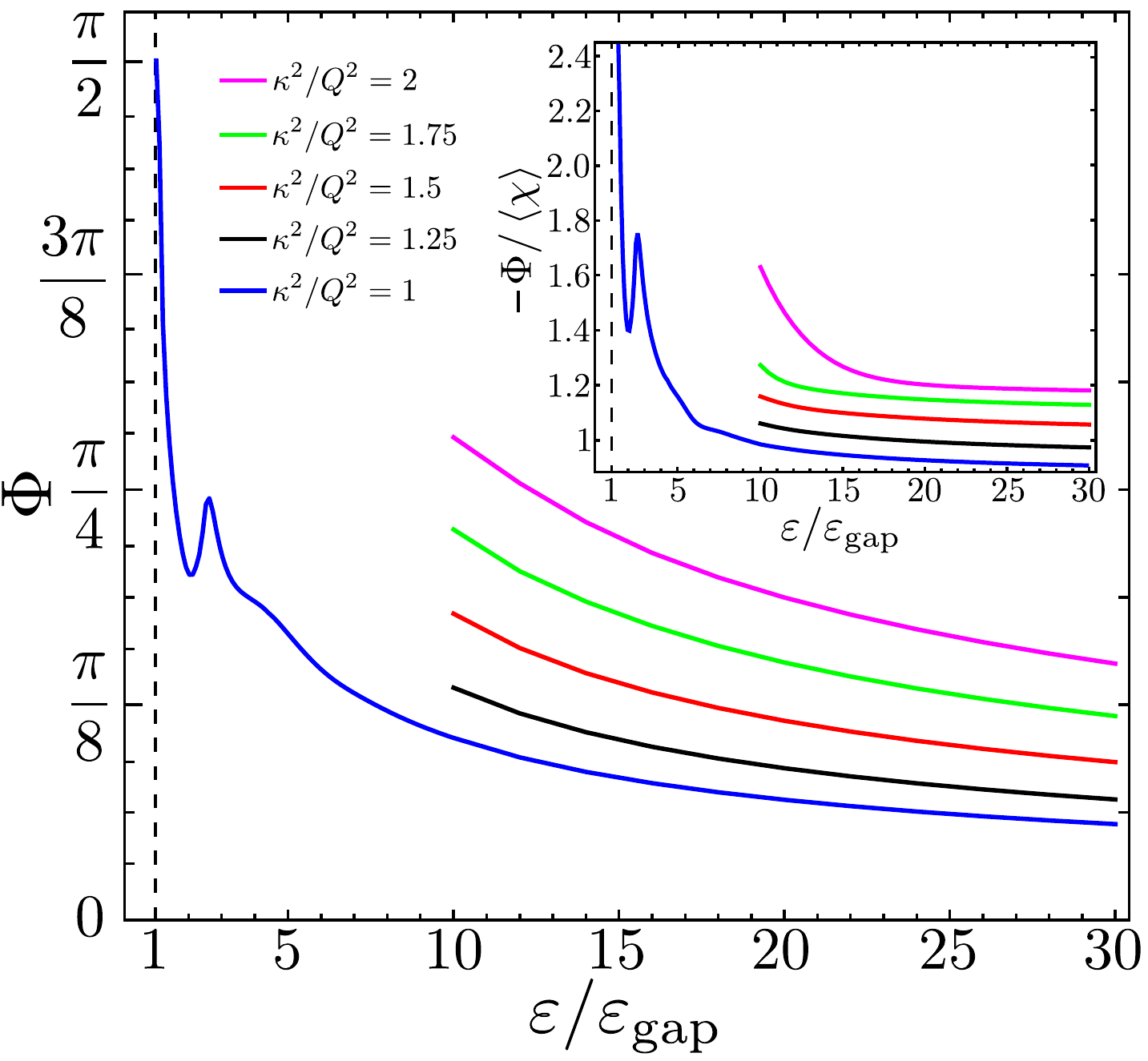}
\caption{The skyrmion Hall angle $\Phi$ as defined in Eq.~\eqref{SkyrmionHallAngle} as a function of energy for various values of $\kappa$, for details see text.
The inset shows the ratio $-\Phi/\langle \chi \rangle_\varepsilon$. The Hall angle is maximal $\Phi= \pi/2$ in the limit $\varepsilon \to \varepsilon_{\rm gap}$ where magnon $s$-wave scattering prevails.
}
\label{fig:SkyrmionHall}
\end{figure}

To be more precise, we define the skyrmion Hall angle, see Fig.~\ref{fig:magnonpressure},
\begin{align} \label{SkyrmionHallAngle}
\Phi \equiv \arctan \frac{d_t \vect R_y}{-d_t \vect R_x} =  
\arctan \frac{\sigma_\parallel(\varepsilon)}{\sigma_\perp(\varepsilon)}.
\end{align}
An evaluation of $\Phi$ as a function of energy for various values of $\kappa$ is shown in Fig.~\ref{fig:SkyrmionHall}.
The Hall angle $\Phi$ for values $\kappa^2/Q^2 = 1.25, 1.5, 1.75$ and $2$ is here evaluated within the WKB approximation and therefore presented only for energies $\varepsilon/\varepsilon_{\rm gap} > 10$. 
The behavior of $\Phi$ for lower energies is shown for $\kappa^2 = Q^2$, for which we calculated the cross sections by using the exact phase shifts for angular momenta $-4 \leq m \leq 3$ while the remaining phase shifts up to $|m| \leq 30$ are again evaluated within the WKB approximation and higher angular momenta are neglected. The scattering at low energies $\varepsilon \to \varepsilon_{\rm gap}$ is governed by $s$-wave scattering corresponding to the phase shift $\delta_m$ with $m=1$ in Eq.~\eqref{ScaAmplitude}, see also Fig.~\ref{fig:phaseshift}. In this limit, the magnon skew scattering is negligible, $\sigma_\perp$ approximately vanishes, and the force \eqref{MagnonPressure} is practically longitudinal to the incoming magnon momentum, $\vect F \propto \hat x$. This results in a maximum Hall angle of $\Phi = \pi/2$ for $\varepsilon \to \varepsilon_{\rm gap}$. For larger energies $\varepsilon/\varepsilon_{\rm gap} \gg 1$, on the other hand, there is substantial magnon skew scattering giving rise to a finite $\sigma_\perp$ and thus a finite transversal force $F_{\perp} \propto \sigma_{\perp}$ that reduces $\Phi$. The peak in the Hall angle at around $\varepsilon/\varepsilon_{\rm gap} \approx 2.5$ for $\kappa^2 = Q^2$ can be attributed to the resonance of the $m=-3$ mode, see Fig.~\ref{fig:phaseshift}.

The inset of Fig.~\ref{fig:SkyrmionHall} compares the skyrmion Hall angle $\Phi$ with the first moment $\langle \chi \rangle_\varepsilon$, with $\langle \mathcal{O}(\chi) \rangle_\varepsilon = \frac{1}{\sigma} \int_0^{2\pi} d\chi \mathcal{O}(\chi) d\sigma/d\chi$. It was suggested in Ref.~\onlinecite{Iwasaki2014} that the ratio $-\Phi/\langle \chi \rangle_\varepsilon$ assumes the value of $1/2$, which we cannot confirm. This discrepancy is probably attributed to an insufficient precision in the 
numerical experiment of Ref.~\onlinecite{Iwasaki2014}.
Inspection of Eq.~\eqref{SkyrmionHallAngle} reveals that a ratio of $1/2$ would only obtain 
for a differential cross section with vanishing higher cumulants so that, e.g., $\langle \sin \chi \rangle_\varepsilon = \sin \langle \chi \rangle_\varepsilon$. In the limit $\varepsilon \to \varepsilon_{\rm gap}$ the ratio $-\Phi/\langle \chi \rangle_\varepsilon$ even diverges as $\langle \chi \rangle_\varepsilon \to 0$ in the limit of $s$-wave scattering.


\section{Summary and discussion}
\label{sec:summary}

A two-dimensional chiral magnet adopts a field-polarized ground state for a sufficiently large magnetic field perpendicular to the film corresponding to $\kappa > \kappa_{\rm cr}$ in Fig.~\ref{fig:skyrmionEnergy}. The magnetic skyrmion then corresponds to a large-amplitude excitation with positive energy $\varepsilon_s$. Such topological skyrmion excitations are always present in an experimental system. Their density might either be given by a thermal distribution reducing to a Boltzmann factor $e^{-\beta \varepsilon_s}$ at lowest temperatures, or depend on the history of the sample because the topological protection of skyrmions results in long equilibration times.

In any case, the presence of a skyrmion will modify the properties of the small amplitude fluctuations, i.e., the magnon excitations. We investigated the basic aspects of this magnon-skyrmion interaction in the present work. Starting from the non-linear sigma model for chiral magnets \eqref{action}, we first derived the magnon Hamiltonian \eqref{Hamiltonian1} by expanding around the skyrmionic saddle point solution. As the skyrmion scattering potential does not preserve the magnon number, this Hamiltonian was found to possess a (bosonic) Bogoliubov-deGennes form. The Hamiltonian also includes a term that explicitly depends on the skyrmion velocity \eqref{MagnonVelocityInteraction}, whose consequences, however, have not been investigated yet in this work

Solving the Bogoliubov-deGennes scattering problem we first determined the magnon spectrum. In the magnetic field regime where the field-polarized state is stable, we found two magnon-skyrmion bound states: the breathing mode with $m=0$ and the quadrupolar mode with $m=-2$, see Figs.~\ref{fig:spectrum} and \ref{fig:BoundStatesVis}. 
For a visualization of these modes see also Ref.~\onlinecite{Movies}. 
For a smaller but finite magnetic field, the quadrupolar eigenfrequency eventually vanishes signalling a local bimeron instability of the single skyrmion.
Our results are in good agreement with a recent numerical diagonalization study.\cite{Lin2014}

The bound magnon modes will give rise to weak subgap resonances in  magnetic or electric resonance experiments whose weight will be proportional to the skyrmion density in the sample. 
In order to determine their selection rules, we consider here the magnetic dipole, $\mathcal{M}_i$, quadrupolar, $\mathcal{Q}_{ij}$, and toroidal moment, $\mathcal{T}_i$,\cite{Spaldin2008} defined as follows 
\begin{align}
\mathcal{M}_i &=  \int d \vect r\,  \hat n_i(\vect r), 
\nn\\
\label{MagneticMoments}
\mathcal{Q}_{ij} &=   \int d \vect r\, (\hat n_i(\vect r) \hat n_j(\vect r) - \frac{1}{3} \delta_{ij}),
\\\nn
\mathcal{T}_i &=   \int d \vect r\, \epsilon_{i \alpha j } \vect r_\alpha \hat n_j(\vect r).
\end{align}
In the field-polarized state without a skyrmion, only the components $\mathcal{M}_3$ and $\mathcal{Q}_{ii}$ with $i=1,2,3$ are finite and proportional to the volume $\mathcal{V}$ whereas all other components vanish. The skyrmion not only leads to a modification of these finite components of order $\mathcal{O}(\mathcal{V}^0)$ but also induces a finite toroidal moment $\mathcal{T}_3$ that scales as the third power of the skyrmion radius, $1/\kappa$, 
\begin{align}
\mathcal{T}_3 = 2\pi \int_0^\infty d\rho \rho^2 \sin\theta = \frac{1}{\kappa^{3}} T(\kappa^2/Q^2),
\end{align}
where $T$ is a dimensionless function. 
If the skyrmion is excited either with the breathing mode or the quadrupolar mode, oscillations of certain moments arise that are listed in Table \ref{tab:MagnonParameters}.
An ac magnetic or electric field could couple to these moments and thus excite the corresponding magnon-skyrmion bound state. An ac magnetic field perpendicular to the film directly couples to $\mathcal{M}_3$, and will therefore excite, similar to the skyrmion lattice,\cite{Mochizuki2012} the breathing mode of the skyrmion. A perpendicular ac electric field, on the other hand, couples to the polarization $\mathcal{P}_3$, which for the insulating chiral magnet Cu$_2$OSeO$_3$ is in general given by\cite{Seki2012-2,YHLiu2013} $\mathcal{P} \propto (\mathcal{Q}_{23},\mathcal{Q}_{31},\mathcal{Q}_{12})$, so that it will excite the quadrupolar mode. As the toroidal moment couples to the cross product of electric and magnetic field, $\mathcal{T} \sim E\times H$,\cite{Spaldin2008} in-plane ac electric and magnetic fields might also be able to excite the breathing mode.

\renewcommand{\arraystretch}{2}
\begin{table}
\begin{tabular}{c|c}
\hline
breathing mode $m=0$ &\, $\mathcal{M}_3, \mathcal{Q}_{11}, \mathcal{Q}_{22}, \mathcal{Q}_{33}, \mathcal{T}_{3}$
\\ 
\hline
quadrupolar mode $m = -2$\, & $\mathcal{Q}_{11}, \mathcal{Q}_{22},  \mathcal{Q}_{12},  \mathcal{Q}_{21}$
\\ 
\hline
\end{tabular}
\caption{Moments that are excited by the magnon-skyrmion bound state $m=0$ and $m=-2$. $\mathcal{M}_i$, $\mathcal{Q}_{ij}$, and 
$\mathcal{T}_{i}$ are the dipole, quadrupolar and toroidal moments, respectively, see Eq.~\eqref{MagneticMoments}.
}
\label{tab:MagnonParameters}
\end{table}

Apart from the spectrum of magnon bound states, we also discussed the differential cross section $d\sigma/d\chi$ for magnon scattering off the skyrmion. 
The scattering potential, in particular, includes an Aharonov-Bohm flux density that governs the scattering characteristics at large magnon energies. While the total flux vanishes, the density possesses a singularity at the skyrmion core that is related to the non-trivial skyrmion topology. As explained in detail in section \ref{sec:DiscScattering}, this specific flux density profile is at the origin of rainbow scattering giving rise to an asymmetric differential cross section with oscillating behavior as a function of the scattering angle, see Fig.~\ref{fig:diffcrosssection}. Our results for $d\sigma/d\chi$ differ from the theory presented in Ref.~\onlinecite{Iwasaki2014}, that did not find oscillations of $d\sigma/d\chi$ characteristic for rainbow scattering. 
Importantly, the magnons skew scatter from the skyrmion, see Fig.~\ref{fig:WKBwavefunction}, as previously pointed out in Ref.~\onlinecite{Iwasaki2014}. This skew scattering generates a topological magnon Hall effect\cite{Mochizuki2014} that is proportional to the skyrmion density.

The magnons not only scatter off but also transfer momentum to the skyrmion. The resulting magnon pressure on the skyrmion due to a constant magnon current was determined in section \ref{sec:MagnonPressure}. Starting from the conservation law associated with translation invariance of the action \eqref{action}, that is discussed in detail in appendix \ref{app:Noether}, 
we demonstrated that the magnon pressure enters as an effective force in the Thiele equation of motion \eqref{MagnonPressureEoM} for the skyrmion. 
This effective force is determined by the net transfer of magnon momentum. It depends, in particular, on the magnon differential cross section, and it is thus a reactive momentum-transfer force.
The solution of the resulting Thiele equation predicts a skyrmion velocity with a component longitudinal and transverse to the magnon current. The transverse component is attributed to the longitudinal scattering cross section $\sigma_\parallel$, giving rise to a large skyrmion Hall effect. 
Due to the asymmetric skew scattering, there is also a finite transversal cross section $\sigma_\perp$, which determines the longitudinal motion. This longitudinal motion is, interestingly, antiparallel to the magnon current, i.e., it is  towards the magnon source. Our theory provides an explanation of the numerical experiment in Ref.~\onlinecite{Iwasaki2014} where a corresponding skyrmion motion was observed. 

We note that our result for the magnon pressure due to momentum-transfer is distinctly different from the conclusion drawn 
from the mode-decomposition theory of Ref.~\onlinecite{Kovalev2012}, which has been invoked to explain the skyrmion motion in Refs.~\onlinecite{Kong2013,Lin2013-3,Kovalev2014}. The latter theory, in particular, predicts in the limit of vanishing Gilbert damping a vanishing skyrmion Hall effect and a universal longitudinal skyrmion motion antiparallel to the magnon current that is independent of the differential magnon scattering cross section in contrast to our findings, Eq.~\eqref{SkyrmionVelocity}. 
We believe that our results 
for the magnon scattering cross section as well as for the magnon pressure 
pave the way for a genuine theory of skyrmion caloritronics,\cite{Bauer2012} i.e., 
thermal spin-transport phenomena for a dilute gas of skyrmions in clean insulating chiral magnets.

There are several interesting open issues to be explored in future work. 
Most importantly, the interaction between the magnons and the skyrmion velocity, Eq.~\eqref{MagnonVelocityInteraction}, naturally produces in second-order perturbation theory an effective retarded skyrmion mass. It is an important open question whether such a retarded mass allows for an additional collective magnon-skyrmion bound state corresponding to a cyclotron mode of the massive Thiele equation.\cite{Ivanov1989,Ivanov2005} Such a mode would be reminiscent of the gyration modes observed as magnetic resonances in the skyrmion lattice phase.\cite{Mochizuki2012,Onose2012,Schwarze2014} 

\acknowledgments

We acknowledge helpful discussions with J.~Iwasaki, S.~Komineas, S.-Z.~Lin, N.~Nagaosa, and, especially, A.~Rosch.
We  would also like to acknowledge an interesting conversation with B.~A.~Ivanov that motivated this work.

\appendix

\section{Conserved Noether currents}
\label{app:Noether}

We present a discussion of important Noether currents of the Lagrangian $\mathcal{L}$ of Eq.~\eqref{action}, i.e., the conservation laws deriving from translation and rotation invariance. Whereas these conservation laws are manifest spin-gauge invariant, the corresponding canonical energy-momentum tensor and angular momentum vector current themselves depend on the spin-gauge potential.\cite{Haldane1986,Volovik1987,Papanicolaou1990,Yan2013} In general, this precludes the definition of a conserved total canonical momentum and angular momentum. However, in case the topological charge is preserved within the sample a definition of a conserved total momentum and angular momentum becomes possible in agreement with previous findings by Papanicolaou and Tomaras.\cite{Papanicolaou1990} For the indices we use the conventions of Table~\ref{tab:Indices}.

\subsection{Canonical energy-momentum tensor}

The theory \eqref{action} is translationally invariant so that the canonical energy momentum tensor
\begin{align}
T_{\mu\nu} &= \frac{\partial \mathcal{L}}{\partial (d_\mu \hat n)} d_\nu \hat n - \delta_{\mu\nu} \mathcal{L}
\end{align}
is conserved
\begin{align} \label{EMConservation}
d_\mu T_{\mu\nu} = 0.
\end{align}
The tensor $T_{\mu\nu}$ depends on the spin-gauge field $\vec A$ of Eq.~\eqref{Ldyn}. For example, the canonical momentum density 
\begin{align}
T_{0\alpha} = -\frac{\hbar}{a^2}\vec A d_\alpha \hat n
\end{align}
is determined by $\vec A$.\cite{Haldane1986} Nevertheless, the conservation law \eqref{EMConservation} itself is gauge invariant. This is best seen by first defining the energy-momentum tensor deriving from the static Lagrangian $\mathcal{L}_{\rm stat}$ only,
\begin{align} \label{EMTensor}
T^{\rm stat}_{\mu\nu} &= \frac{\partial \mathcal{L}_{\rm stat}}{\partial (d_\mu \hat n)} d_\nu \hat n - \delta_{\mu\nu} \mathcal{L}_{\rm stat}.
\end{align}
%
%
The conservation law \eqref{EMConservation} can then be written in the manifest spin-gauge invariant form
\begin{align} \label{EMConservation1}
d_\mu T^{\rm stat}_{\mu\nu} - \frac{4\pi\hbar}{a^2} \epsilon_{\mu 0 \nu} j_\mu^{\rm top} = 0,
\end{align}
where we have used the identity of Eq.~\eqref{TopCurrentSpinGauge}.
So we arrive at the result that the divergence of $T^{\rm stat}_{\mu\nu}$ is given by the topological current of Eq.~\eqref{TopCurrent}.

Assuming that the integral over the spatial divergence $\int d^2 \vect r d_\alpha T^{\rm stat}_{\alpha \mu} = 0$, i.e., that $T^{\rm stat}_{\alpha \mu}$ vanishes on the surface of the two-dimensional sample, it follows from the time-component of Eq.~\eqref{EMConservation1} the conservation of total energy, $d_0 E = 0$, with
\begin{align}
E \equiv - \int d^2 \vect r T^{\rm stat}_{00} = \int d^2 \vect r \mathcal{L}_{\rm stat}.
\end{align}
Due to the anomalous form of the conservation law, however, we are in general not able to define a conserved total momentum. Instead it follows from the spatial component of \eqref{EMConservation1} with $T^{\rm stat}_{0 \alpha} = 0$ that the spatial integral over the topological current vanishes 
\begin{align} \label{IntTopCurrent}
- \frac{4\pi\hbar}{a^2} \epsilon_{\alpha 0 \beta} \int d^2 \vect r\,  j_\alpha^{\rm top} = 0.
\end{align}
However, with the additional assumption that the conservation law for the topological current of Eq.~\eqref{ConservedTopCurrent} holds and that the product $\vect r_\alpha j_\beta^{\rm top}$ vanishes on the surface, we can write
\begin{align}
&\int d^2 \vect r j_\alpha^{\rm top} = \int d^2 \vect r \left(- d_\beta (\vect r_\alpha j_\beta^{\rm top}) + j_\alpha^{\rm top} \right)
\\\nn
&= - \int d^2 \vect r\, \vect r_\alpha  d_\beta  j_\beta^{\rm top} = d_0 \int d^2 \vect r\, \vect r_\alpha  j_0^{\rm top}.
\end{align}
In this case, following Ref.~\onlinecite{Papanicolaou1990} we can identify a total conserved momentum 
\begin{align} \label{ConservedMomentum}
\vect P_\alpha =  - \frac{4\pi\hbar}{a^2} \epsilon_{0 \alpha \beta} \int d^2 \vect r\, \,  \vect r_\beta  j_0^{\rm top}
\end{align}
with the first moment of the topological charge distribution. 

In Ref.~\onlinecite{Papanicolaou1990} this result for the conserved momentum was obtained by starting directly from the equation for the time-derivative of the topological charge density
\begin{align}  
\frac{4\pi\hbar}{a^2} d_0 j_0^{\rm top} = 
\epsilon_{0 \alpha \beta} d_\alpha d_\gamma T^{\rm stat}_{\gamma \beta},
\end{align}
which can be derived from Eq.~\eqref{EMConservation1} by applying the derivative $\epsilon_{\alpha 0 \nu} d_\alpha$ with $\alpha = 1,2$, summing over $\nu$, and using the conservation law for the topological current Eq.~\eqref{ConservedTopCurrent}.

\subsection{Angular momentum}

For completeness, we also discuss the conservation law deriving from rotational invariance. Due to the spin-orbit coupling $Q$, only the total angular momentum vector current obeys a conservation law. In the following, we first discuss the spin-, afterwards the orbital and, finally, the total angular momentum. 

\paragraph{Spin angular momentum}

From the rotations by an infinitesimal angle $\omega$ of the magnetisation around the magnetic field direction $\hat B = \hat z$, $\hat n_i \to \hat n_i + \delta \hat n_i$ with
\begin{align}
\delta \hat n_i = \omega \epsilon_{i z k} \hat n_k
\end{align}
follows
\begin{align} \label{SpinAngularMomentum}
d_\mu \mathcal{S}_\mu &= 
- \frac{\partial \mathcal{L}_{\rm stat}}{\partial n_i} \epsilon_{i z k} \hat n_k -
\frac{\partial \mathcal{L}_{\rm stat}}{ \partial d_\mu \hat n_i} \epsilon_{i z k} d_\mu \hat n_k
\\\nn
&= -\varepsilon_0 Q ( \hat n_z d_\alpha \hat n_\alpha - \hat n_\alpha d_\alpha \hat n_z ).
\end{align}
The 2+1 spin angular momentum current,
\begin{align}
\mathcal{S}_0 &= -\frac{\hbar}{a^2} \hat n_z,
\\
\mathcal{S}_\alpha &= - \frac{\partial \mathcal{L}_{\rm stat}}{\partial d_\alpha \hat n_i} \epsilon_{i z k} \hat n_k
\\\nn
&= - \varepsilon_0 [ \epsilon_{izk} \hat n_k d_\alpha \hat n_i + Q (\hat n_z \hat n_\alpha - \delta_{z\alpha})]
\end{align}
is not conserved as the right-hand side of Eq.~\eqref{SpinAngularMomentum} is finite due to the spin-orbit coupling $Q$.
Note that according to our sign convention in Eq.~\eqref{Ldyn} the magnetisation vector points antiparallel to the spin vector as it is for example the case for electrons.

\paragraph{Orbital angular momentum}

Performing an infinitesimal orbital rotation around the $z$-axis
\begin{align}
\delta \hat n_i = - \omega \epsilon_{\alpha 0 \beta} \vect r_{\beta} d_\alpha \hat n_i 
\end{align} 
we obtain 
\begin{align} \label{OrbitalAngularMomentum}
d_\mu L_\mu = \epsilon_{\beta 0 \alpha} T_{\alpha \beta}.
\end{align}
The angular momentum 
\begin{align}
L_\mu = \epsilon_{\beta 0 \gamma} \vect r_{\gamma} T_{\mu \beta}
\end{align}
is not conserved because the energy-momentum tensor is not symmetric due to the spin-orbit coupling $Q$.

\paragraph{Total angular momentum}

The theory \eqref{action} is invariant with respect to a combined rotation of spin and real space around the magnetic field direction. As a result, the total angular momentum
\begin{align}
J_\mu = \mathcal{S}_\mu + L_\mu
\end{align}
is conserved
\begin{align} \label{AngularMomentum}
d_\mu J_\mu = 0,
\end{align}
which follows from Eqs.~\eqref{SpinAngularMomentum} and \eqref{OrbitalAngularMomentum}.
Similarly as for the energy-momentum tensor, the conservation law \eqref{AngularMomentum} is spin-gauge invariant whereas $J_\mu$ itself is not. Introducing the part of the angular momentum current
\begin{align}
\tilde J_\mu = \mathcal{S}_\mu + \epsilon_{\beta 0 \gamma} \vect r_{\gamma} T^{\rm stat}_{\mu \beta}
\end{align}
that is manifest invariant, the conservation law assumes the form
\begin{align}
d_\mu \tilde J_\mu + \frac{4\pi\hbar}{a^2} \vect r_\alpha j_\alpha^{\rm top} = 0.
\end{align}

With the assumption that the current $ \tilde J_\alpha$ vanishes on the surface of the two-dimensional sample $\int d^2 \vect r d_\alpha \tilde J_\alpha = 0$ we find
\begin{align}
d_0 \int d^2 \vect r\,  \tilde J_0 + \frac{4\pi\hbar}{a^2}  \int d^2 \vect r\,  
\vect r_\alpha j_\alpha^{\rm top} = 0.
\end{align}
The time-dependence of the spatial integral over the density, $\tilde J_0 = -\frac{\hbar}{a^2} \hat n_z$, is thus determined by the integrated scalar product of the position vector $\vect r$ with the spatial topological current $j_\alpha^{\rm top}$. 
With the additional assumption that the topological current is conserved, Eq.~\eqref{ConservedTopCurrent}, and that 
the product $\vect r^2 j_\alpha^{\rm top}$ vanishes on the surface, we can rewrite the last term as a time derivative
\begin{align}
&\int d^2 \vect r\, \vect r_\alpha j_\alpha^{\rm top} = \int d^2 \vect r\, \frac{1}{2}\Big(d_\alpha(\vect r^2 j_\alpha^{\rm top}) - \vect r^2 d_\alpha j_\alpha^{\rm top}\Big)
\nn\\&
= d_0 \int d^2 \vect r\, \frac{1}{2} \vect r^2 j_0^{\rm top}.
\end{align}
As a result, we finally obtain
\begin{align} \label{ConservedAngularMomentum}
\mathcal{J} = \int d^2 \vect r\,  \left( -\frac{\hbar}{a^2} \hat n_z + \frac{2\pi \hbar}{a^2} \vect r^2 j_0^{\rm top} \right)
\end{align}
for the conserved total angular momentum, $d_0 \mathcal{J} = 0$.\cite{Papanicolaou1990}

\subsection{Momentum and angular momentum of the skyrmion}
\label{app:Mom&AngMom}

It is instructive to compute the momentum and the angular momentum attributed to the magnetic skyrmion solution of section \ref{sec:saddlepointskyrmion}. Neglecting the massive fluctuations, $\psi$, we obtain for the momentum \eqref{ConservedMomentum}
\begin{align}
\vect P^{\rm skyr}_\alpha &= - \frac{4\pi\hbar}{a^2} \epsilon_{0 \alpha \beta} \int d^2 \vect r\, \,  \vect r_\beta  j_0^{\rm top}(\vect r - \vect R)\Big|_{\psi,\psi^*=0}
\nn\\&= \frac{4\pi\hbar}{a^2} \epsilon_{0 \alpha \beta} \vect R_\beta. 
\end{align}
On this level of approximation, the conservation of $\vect P^{\rm skyr}_\alpha$ directly follows from the equation of motion \eqref{ClassicalEoM},
\begin{align} \label{ClassicalEoM-MomentumFlow}
d_0 \vect P_\alpha^{\rm skyr} = \frac{4\pi\hbar}{a^2} \epsilon_{0 \alpha \beta} d_0 \vect R_\beta = (\vect G \times d_0 \vect R)_\alpha =  0
\end{align}

The angular momentum \eqref{ConservedAngularMomentum} of the fully field-polarized state, $\hat n = \hat z$, is already non-zero and given by the total spin of the sample, $\mathcal{J}_{\rm FP} = - \mathcal{V}\hbar/a^2$ where $\mathcal{V}$ is the volume.
Neglecting again the massive fluctuations, we find for the angular momentum attributed to the skyrmion
\begin{align}
\mathcal{J}_{\rm skyr} &= \mathcal{J}\Big|_{\psi,\psi^* = 0} - \mathcal{J}_{\rm FP}
\\\nn&=
\int d^2 \vect r\,  \left( -\frac{\hbar}{a^2} (\hat n_z - 1) + \frac{2\pi\hbar}{a^2} \vect r^2 j_0^{\rm top} \right)\Big|_{\psi,\psi^* = 0}
\\\nn&= 
\int d^2 \vect r\,  \left( -\frac{\hbar}{a^2} (\hat n_z - 1) + \frac{2\pi\hbar}{a^2} (\vect r-\vect R)^2 j_0^{\rm top} \right)\Big|_{\psi,\psi^* = 0}
\\\nn &\quad+\int d^2 \vect r\,  \left( \frac{2\pi\hbar}{a^2} (2 \vect r \vect R- \vect R^2) j_0^{\rm top} \right)\Big|_{\psi,\psi^* = 0}
\end{align}
It turns out that the first line in the last equation exactly vanishes,
\begin{align}
&\int d^2 \vect r\,  \left( -\frac{\hbar}{a^2} (\hat n_z - 1) + \frac{2\pi\hbar}{a^2} (\vect r-\vect R)^2 j_0^{\rm top} \right)\Big|_{\psi,\psi^* = 0}\nn
\\\nn&= \frac{2\pi\hbar}{a^2} \int_0^\infty d\rho\,\rho  \left( -(\cos \theta - 1) + \frac{\rho}{2} \theta' \sin\theta\right) 
\\&= \frac{2\pi\hbar}{a^2} \rho^2 \sin^2 \frac{\theta}{2} \Big|_{\rho =0}^\infty = 0,
\end{align}
due to the exponentially fast approach  of the polar angle $\theta$ to its boundary value at large distances, $\theta \to 0$. The change of spin angular momentum due to the spatial dependence of $\hat n_z(\vect r)$ is thus exactly compensated by the orbital angular momentum that is carried by the skyrmion texture. The skyrmion angular momentum then reduces to 
\begin{align}
\mathcal{J}_{\rm skyr} &= - \frac{2\pi\hbar}{a^2}   \int d^2 \vect r\,  (2 \vect r \vect R- \vect R^2) j_0^{\rm top} \Big|_{\psi,\psi^* = 0} 
\nn\\&= -\frac{2\pi\hbar}{a^2} \vect R^2.
\end{align}
The time derivative of $\mathcal{J}_{\rm skyr}$, i.e., the torque assumes with the help of Eq.~\eqref{ClassicalEoM-MomentumFlow} the following intuitive form
\begin{align}
d_0 \mathcal{J}_{\rm skyr} = (\vect R \times d_0 \vect P_{\rm skyr})_z.
\end{align}
%



\begin{thebibliography}{}
\bibitem{RemarkUmklapp}
We assume that Umklapp scattering can be neglected as the size of the skyrmion, $1/\kappa$, is incommensurate and large compared to the lattice spacing $a$.
\end{thebibliography}
\end{document}